\documentclass[aps,prl,twocolumn]{revtex4-2}

\usepackage{graphicx}
\usepackage{dcolumn}
\usepackage{bm}
\usepackage{amsmath}
\usepackage{amssymb}
\usepackage{url}

\usepackage{color}
\usepackage[colorlinks,bookmarks=false,citecolor=darkblue,linkcolor=red,urlcolor=blue]{hyperref}

\definecolor{darkred}{rgb}{0.7,0.0,0.0}
\definecolor{darkblue}{rgb}{0,0.02,0.45}

\graphicspath{{figures/}}

\begin{document}

\title{Self-reconstruction of order parameter in spin-triplet superconductor UTe$_2$}

\author{Y. Tokiwa$^{1,\star}$}
\author{P. Opletal$^1$}
\author{H. Sakai$^1$}
\author{K. Kubo$^1$}
\author{S. Kambe$^1$}
\author{E. Yamamoto$^1$}
\author{M. Kimata$^2$}
\author{S. Awaji$^2$}
\author{T. Sasaki$^2$}
\author{D. Aoki$^3$}
\author{Y. Yanase$^4$}
\author{Y. Tokunaga$^1$}
\author{Y. Haga$^1$}

\affiliation{$^1$Advanced Science Research Center, Japan Atomic Energy Agency, Tokai, Ibaraki 319-1195, Japan}
\affiliation{$^2$Institute for Materials Research, Tohoku University, Sendai, Miyagi 980-8577, Japan.}
\affiliation{$^3$Institute for Materials Research, Tohoku University, Oarai, Ibaraki 311-1313, Japan.}
\affiliation{$^4$Department of Physics, Kyoto University, Kyoto 606-8502, Japan}
\affiliation{{\rm Correspondence and requests for materials should be addressed to Y.T. (email: yoshifumi.tokiwa@jaea.go.jp)}}

\begin{abstract}

We investigate the effect of easy-axis metamagnetic crossover on superconductivity in UTe$_2$ along the $a$-axis through measurements of AC susceptibility, magnetization, and the magnetocaloric effect. In ultra-clean single crystals, we identify a field-induced phase transition within the superconducting state at 5.6 T, driven by metamagnetism. This transition leads to a high-field superconducting state, significantly increasing the upper critical field to 12 T. A sudden increase in entropy at the transition suggests a self-reconstruction of the order parameter, enabling multi-component superconducting states to adapt to external perturbations.

\end{abstract}

\maketitle

Unconventional superconductivity breaks not only the U(1) gauge symmetry but also additional symmetries of its normal-conducting state. The symmetry of a normal-conducting state can be broken in various ways, making it possible for different superconducting (SC) states to be realized. Observations of phase transitions within the superconductivity of several materials, such as UPt$_3$ and CeRh$_2$As$_2$, are evidence for multiple SC states\cite{Joynt2002,Khim2021}. Additionally, external symmetry-breaking influences such as magnetic field or uniaxial stress can render these SC states indistinguishable. This causes intermixing of order parameters, leading to multiple-component superconductivity. Mixing components with different orbital and spin configurations enables the optimization of superconductivity in response to external perturbations.

The recently discovered superconductor UTe$_2$ ($T_c$=1.6 K) \cite{Ran2019} attracts much attention due to its unconventional SC properties, which suggest spin-triplet pairing.  Multiple SC phases have been observed under pressure and in magnetic fields applied along the $b$-axis, which is perpendicular to the easy magnetic $a$-axis~\cite{Aoki2020a,Rosuel2022,Sakai2023,Kinjo2022,Braithwaite2019}. For this magnetic field direction $T_c$ is initially suppressed but, remarkably, experiences enhancement above 15 T \cite{Knafo2019,Knafo2021,Rosuel2022}. Moreover, at specific field angles between the $b$ and $c$-axes, the superconductivity is entirely suppressed but reappears at a very high field of 40 T \cite{Ran2019a,Knebel2019}.

Significant efforts have also been devoted to improving the crystal quality of UTe$_2$~\cite{Haga2022,Rosa2022a,Aoki2022a,Sakai2022}. Recently, we developed the molten salt flux (MSF) method as a novel approach to grow ultraclean UTe$_2$ crystals~\cite{Sakai2022}. This method has led to a substantial increase in $T_c$, exceeding 2 K, compared to the initially reported $T_c$ of 1.6 K for crystals grown using the chemical vapor transport (CVT) method ~\cite{Rosa2022a,Aoki2022a,Sakai2022,Aoki2022c}.

Despite intensive research, the symmetry of the SC order parameter in UTe$_2$ is still under extensive debate~\cite{Jiao2020,Kittaka2020,Xu2019,Ishizuka2019,Machida2021,Ishihara2023,Kanasugi2022,Lee2023_arxiv,Ajeesh2023,Suetsugu2023,Matsumura2023,Lee2023_arxiv,Hayes2024,Theuss2024,Gu,Tei}.
Within the orthorhombic $D_{2h}$ point group at zero field, an odd-parity order parameter is classified into one of four irreducible representations: $A_{\rm u}$, $B_{\rm 1u}$, $B_{\rm 2u}$, and $B_{\rm 3u}$. The $A_{\rm u}$ symmetry is fully gapped, whereas the other symmetries, $B_{\rm 1u}$, $B_{\rm 2u}$, and $B_{\rm 3u}$, feature point nodes on the Fermi surface along the $c$-, $b$-, and $a$-axes, respectively.
Applying a magnetic field further reduces symmetry and induces intermixing of the zero-field order parameters. For instance, when a magnetic field is applied along the $a$-axis, the symmetry decreases from $D_{\rm 2h}$ to $C_{\rm 2h}$. This reduction renders the representations $A_{\rm u}$ and $B_{\rm 3u}$, as well as $B_{\rm 1u}$ and $B_{\rm 2u}$, indistinguishable, leading to SC states of $A_{\rm u}^{\rm 2h} = A_{\rm u} + B_{\rm 3u}$ and $B_{\rm u}^{\rm 2h} = B_{\rm 1u} + B_{\rm 2u}$~\cite{Ishizuka2019}. This intermixing introduces flexibility in the relative weights of each component, a characteristic feature of multi-component SC states.

In this Letter, we report on the magnetic field dependence of entropy ($S$), which reflects quasi-particle excitations, as determined from specific heat ($C$) and magnetocaloric effect (MCE, $(\partial T/\partial B)_S$) measurements. Particular focus is placed on the magnetic Gr\"{u}neisen parameter ($\Gamma_B = (\partial T / \partial B)/T$), which is the temperature coefficient of the MCE and is related to the magnetic field derivative of $S$, expressed as $\Gamma_B = -(\partial S / \partial B)/C$. This makes $\Gamma_B$ a more sensitive probe for detecting magnetic-field-induced phase transitions and crossovers, compared to specific heat, which reflects only the temperature derivative of $S$. Through our measurements, we identified a field-induced phase transition within the SC state for $B \parallel a$-axis. This transition, driven by a metamagnetic (MM) crossover, significantly enhances the upper critical field ($B_{\rm c2}$) in high-quality single crystals. A sharp increase in $S$ at the transition indicates that self-reconstruction of the order parameters occurs in the multi-component SC state, enabling the superconductivity to adapt to the enhanced internal fields induced by the metamagnetism.

\begin{figure}
\includegraphics[width=\linewidth,keepaspectratio]{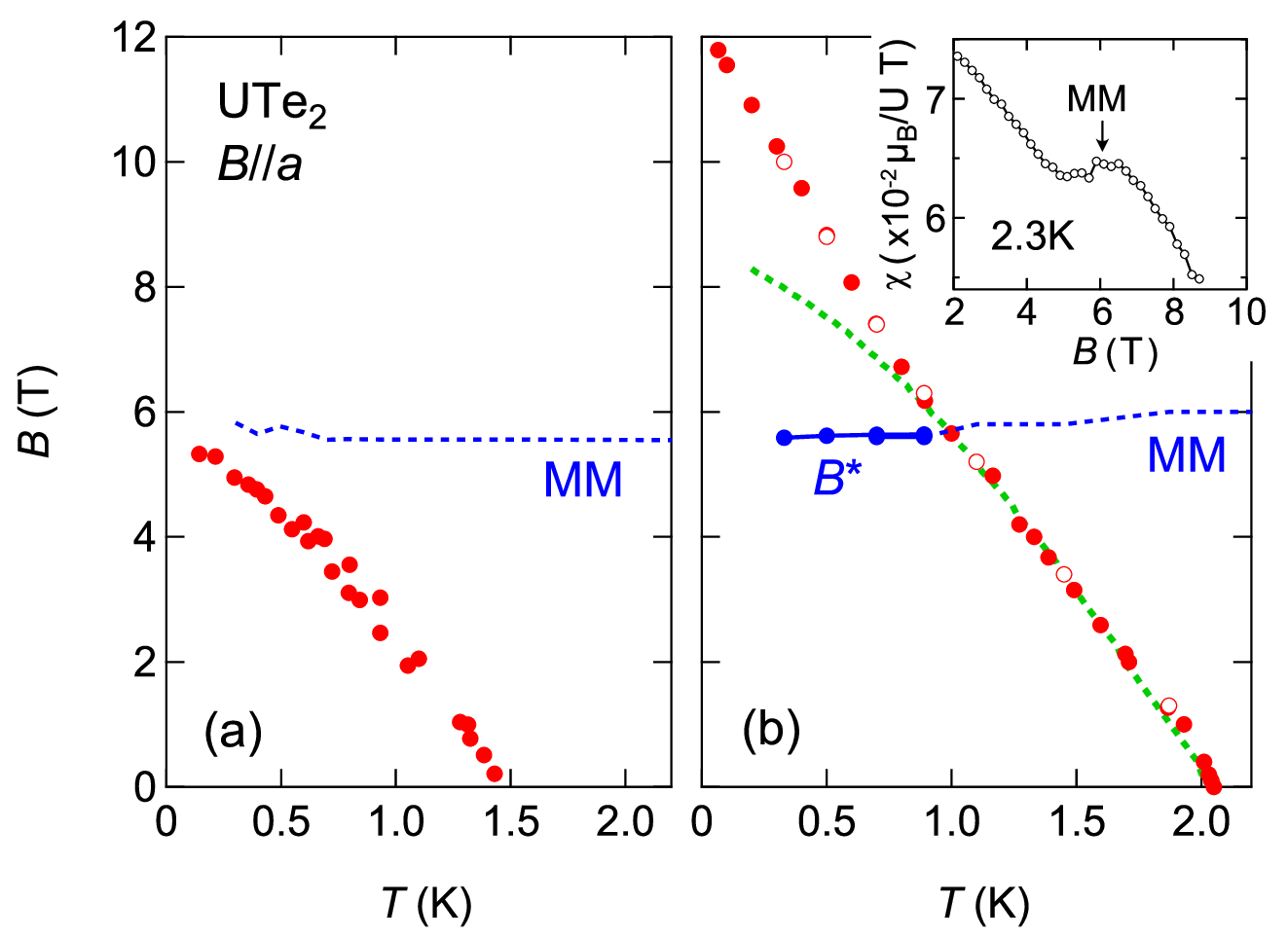}%
\caption{\label{phase} Superconduting phase diagrams of UTe$_2$ samples grown by chemical vapor transport (a)\cite{Niu2020} and molten-salt flux (b) methods for magnetic fields along the $a$-axis. The upper critical field of superconductivity, $B_{\rm c2}$, for samples grown by molten-salt flux method is determined by the AC magnetic susceptibility (solid red circles, see Supplementary Fig. S1 \cite{SM}) and the magnetic Gr\"{u}neisen parameter, $\Gamma_{\rm B}$ (open red circles,\ Supplementary Fig. S3 \cite{SM}). The dotted blue line is the position of broad minimum in $\Gamma_{\rm B}$ due to the metamagnetic crossover (Fig. \ref{MCE}(b)). The solid blue line indicates the first-order phase transition at $B^{\star}$=5.6 T within the superconducting (SC) state, induced by the metamagnetic crossover (Figs. \ref{MCE}(d) and \ref{entropy}(a)). The $B_{\rm c2}$ curve (green dotted line) is obtained by solving the Eliashberg equations in the clean limit and incorporating strong coupling effects through an adjusted strong coupling parameter $\lambda$ to reproduce the initial slope of $B_{c2}$($T$) \cite{Carbotte,Wu2017}. The Pauli-limiting effect is omitted to mimic spin-triplet pairing. Inset shows magnetic susceptibility $\chi$ of UTe$_2$ at 2.3 K as a function of magnetic field along the $a$-direction.}
\end{figure}

The AC magnetic susceptibility was measured at temperatures down to 75 mK and magnetic fields up to 15 T in a $^3$He-$^4$He dilution refrigerator with an SC magnet (see Supplementary Fig. S1 \cite{SM}). Magnetization was measured for the fields up to 9 T using a vibration-sample magnetometer (VSM) in a commercial Physical Property Measurement System of Quantum Design. MCE, defined as $(\partial T/\partial B)_S$, was measured using the alternating-field method as detailed in
Ref. \onlinecite{Tokiwa2011}. High-quality single crystals grown by the MSF method were used. We confirmed the SC transition at $T_c = 2.05$ K for the samples through specific heat measurements at zero field.

Figure \ref{phase}(a) shows the SC phase diagrams for crystals grown using the CVT method~\cite{Niu2020} under magnetic fields applied along the easy $a$-axis. For this field orientation, a weak MM crossover has been reported around 6 T~\cite{Miyake2019,Niu2020}. As the $B_{\rm c2}$ of the low-$T_c$ samples falls short of reaching the MM field, the SC phase boundary does not intersect with the MM crossover line.

The MM field remains unaffected by sample quality. The inset of Fig.~\ref{phase}(b) displays the magnetic susceptibility measured in an MSF crystal as a function of the magnetic field. A maximum of around 6 T indicates the MM crossover, which leads to a stepwise increase in magnetization and, consequently, a sudden enhancement of the internal magnetic field. In MSF-grown high-quality crystals, however, an intriguing interplay between metamagnetism and superconductivity arises: $B_{\rm c2}$ surpasses the MM field, causing the SC phase boundary to intersect with the MM crossover line. As a result, $B_{\rm c2}(T)$ exhibits a kink at the MM field, deviating from the theoretical curve (dotted green line in Fig.~\ref{phase}(b)). This leads to a clear change in the slope of  $B_{\rm c2}(T)$ observed at $B^\star$ (See Supplementary Fig. S2 \cite{SM}). This behavior contrasts sharply with the downward curvature observed in lower $T_c$ samples grown by the CVT method (Fig.~\ref{phase}(a)). This interplay results in a significant enhancement of $B_{\rm c2} = 12$ T, doubling the initial $B_{\rm c2} = 6$ T, despite only a modest 25\% increase in $T_c(0)$ from 1.6 K to 2.0 K~\cite{Ran2019}.

Figures \ref{MCE}(a,b) display the specific heat divided by temperature ($C/T$) and $\Gamma_{\rm B}$ in the normal-conducting state at 2.3 K. In paramagnetic states, $S$ generally decreases with increasing magnetic field as magnetic moments align, resulting in negative $\partial S / \partial B$ and, consequently, positive $\Gamma_B = -(\partial S / \partial B)/C$. However, in the field range around 6 T, where $\Gamma_B$ becomes negative (green shaded region), $S$ increases with magnetic field. This behavior reflects enhanced magnetic fluctuations associated with the MM crossover, a well-established phenomenon supported by experimental studies~\cite{Zacharias13,Tokiwa2013,Rost-Science09}.

Surprisingly, within the SC state, we observed a pronounced anomaly in $\Gamma_B$ at $B^\star = 5.6$ T (Fig.~\ref{MCE}(d)). This sharp anomaly provides strong evidence of a field-induced phase transition within the SC state. Additionally, a step-like increase in $\Gamma_B$ is observed at $B_{\rm c2}=8.6$ T. In contrast,  $C/T$ exhibits only a weak inflection point at $B^\star$ and a peak at $B_{\rm c2}$ (Fig.~\ref{MCE}(c)).

\begin{figure}
\includegraphics[width=\linewidth,keepaspectratio]{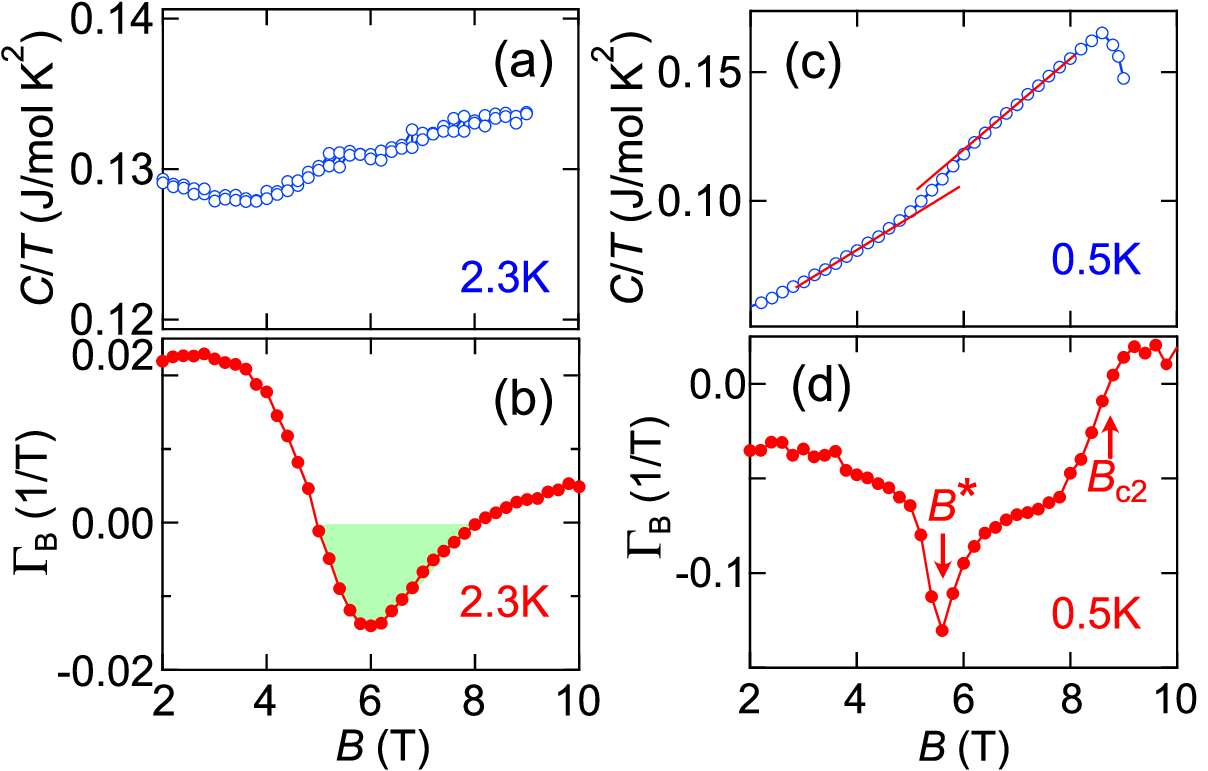}%
\caption{\label{MCE} Specific heat devided by temperature and magnetic Gr\"{u}neisen parameter as functions of magnetic field in the normal-conducting state at 2.3 K (a) and in the superconducting state at 0.5 K (b). The green shaded area in (b) highlights the region where $\Gamma_B$ is negative due to the metamagnetic crossover in the normal-conducting state. The red lines in (c) are guide for eyes.}
\end{figure}

\begin{figure}
\includegraphics[width=\linewidth,keepaspectratio]{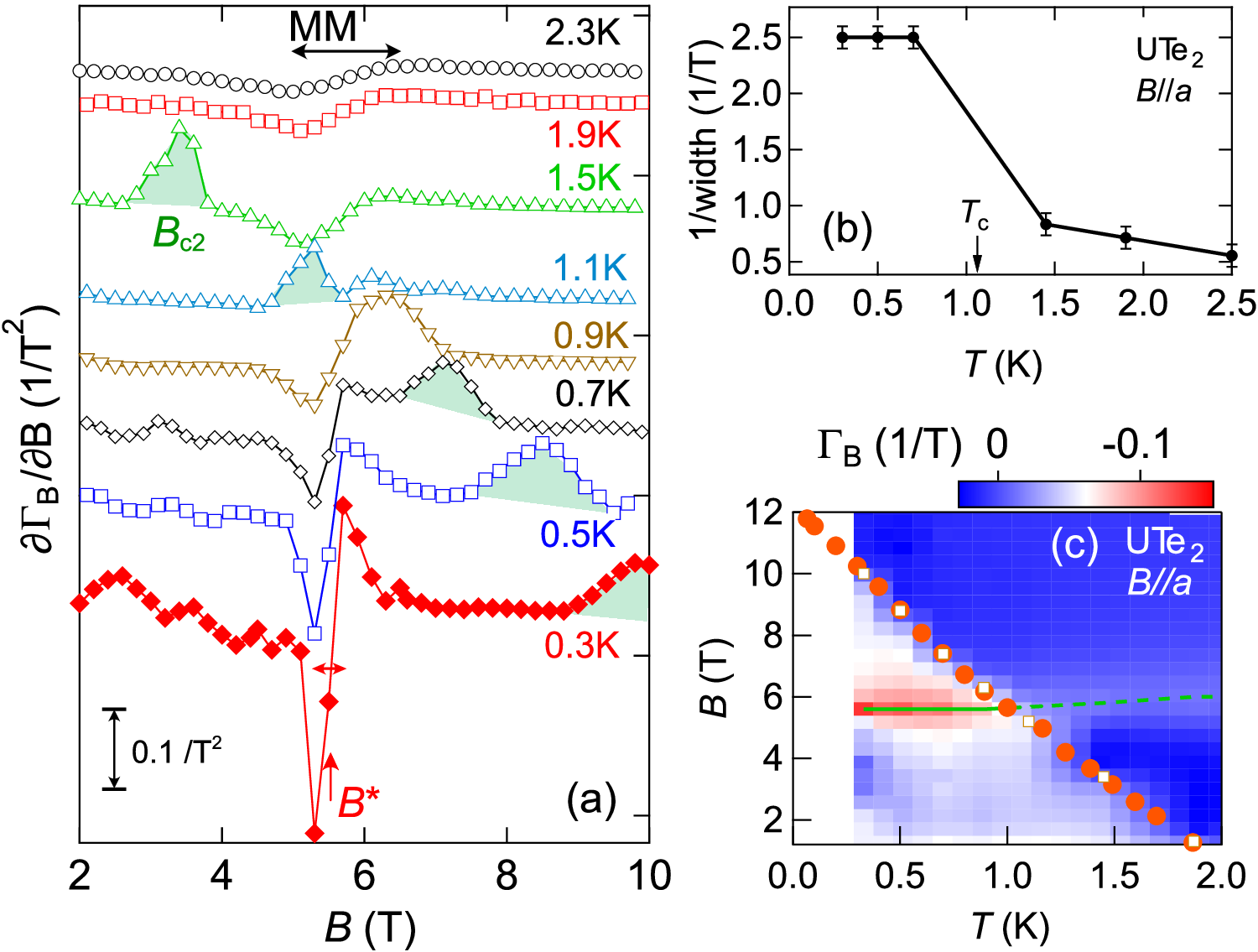}%
\caption{\label{dGdB} (a) Magnetic field derivative of the magnetic Gr\"{u}neisen parameter ($\partial \Gamma_{\rm B}/\partial B$) of UTe$_2$ plotted as a function of magnetic field applied parallel to the $a$-axis at various temperatures. Data sets for different temperatures are vertically shifted for clarity. Green shaded area highlights anomaly of the upper critical field of superconductivity ($B_{\rm c2}$). Double sided arrows indicate the width of the anomaly of metamagnetism and the induced transition inside the superconductivity. (b) Inverse of the width of anomaly at the metamagnetism in the normal-conducting state or at the induced phase transition in the superconducting state. The width is defined as the distance from the minimum and the maximum in $\partial \Gamma_{\rm B}/\partial B$, as indicated by double-sided arrows. The data points at 1.1 and 0.9K are omitted because it is difficult to estimate the width due to the overlapping anomalies of $B_{\rm c2}$ and metamagnetism/induced phase transition. (c) Superconduting phase diagram of UTe$_2$ for magnetic fields along the $a$-axis. The color contour of $\Gamma_{\rm B}$ is added to the phase diagram shown in Fig. \ref{phase}(b) ( Figs.~\ref{MCE}(b,d) and Supplementary Fig.S3 \cite{SM}).  }
\end{figure}

Next, we show that the MM crossover evolves into the phase transition at $B^\star$, resulting in the coincidence between the crossover field and $B^\star$.
Figure \ref{dGdB}(a) shows the magnetic field derivative of the magnetic Gr\"{u}neisen parameter, \( \partial \Gamma_B / \partial B \), as a function of magnetic field at various temperatures. (see Supplementary Fig.S3 for the raw $\Gamma_B$ data \cite{SM}.) In the normal-conducting state, \( \Gamma_B \) exhibits a single minimum at the MM crossover, resulting in a minimum and a maximum in \( \partial \Gamma_B / \partial B \) across the MM crossover, as indicated by the double-sided arrow for the data at 2.3 K. As the temperature decreases, the anomaly at \( B_{\rm c2} \) emerges as a pronounced enhancement, highlighted by the green shading. The value of \( B_{\rm c2} \) increases with decreasing temperature, while the position of the MM anomaly remains nearly unaffected. Interestingly, the MM anomaly becomes significantly sharper upon entering the SC state. In Fig. \ref{dGdB}(b), we plot the inverse width of the anomaly as a function of temperature, where the width is defined as the distance between the minimum and maximum in \( \partial \Gamma_B / \partial B \), as marked by the double-sided arrows. We also indicate \( T_c \) at \( B^\star = 5.6 \, \mathrm{T} \). Notably, the anomaly narrows dramatically across \( T_c \). The \( \Gamma_B \) values are represented in the color contour of the phase diagram shown in Fig. \ref{dGdB}(c), effectively illustrating the transformation from the MM crossover in the normal-conducting state to an induced phase transition within the SC state.

\begin{figure}
\includegraphics[width=\linewidth,keepaspectratio]{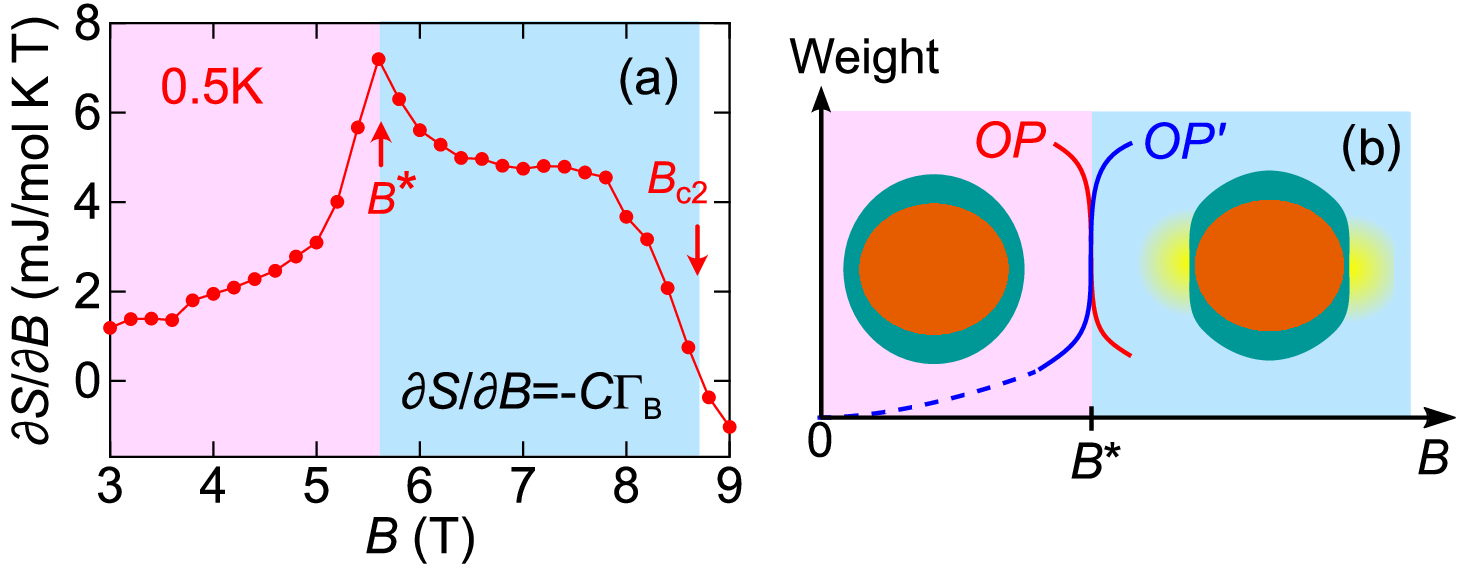}%
\caption{\label{entropy} (a) Magnetic field derivative of entropy $\partial S/\partial B$ of UTe$_2$ as a function of magnetic field along the $a$-axis at temperature of 0.5 K. (b) Schematic behaviour of the weights of superconducting components as a function of magnetic field. See the main text for the definition of the weights. The left/right insets illustrate the SC gap anisotropy of $OP$/$OP'$-dominating low/high-field superconducting states. The orange and green colors indicate Fermi surface (FS) and SC gap ($\Delta$), respectively. The radial gradient yellow colour in the right inset indicates quasi-particle excitations due to the gap minimum caused by the $OP'$ component.}
\end{figure}

We have identified the phase transition and now discuss its nature within the SC state. Figure \ref{entropy}(a) presents the magnetic field derivative of $S$, calculated using $C$ and $\Gamma_{\rm B}$, as \( \partial S / \partial B = -C \Gamma_{\rm B} \). Within the SC state, \( \partial S / \partial B \) exhibits a clear anomaly, as shown in Fig. \ref{entropy}(a), indicative of a phase transition. It is important to note that $S$ exhibits a step at a first-order phase transition, corresponding to a latent heat, whereas it shows a kink at a second-order phase transition. Consequently, $\partial S/\partial B$ displays a peak and a step for first- and second-order phase transitions. In our observations, \( \partial S / \partial B \) reveals a peak and a step-wise decrease  at $B^\star$  and \( B_{\rm c2} \), respectively. This indicates that the induced phase transition at $B^\star$ within the SC state is of first-order, while the transition at \( B_{\rm c2} \) is of second-order. Furthermore, since \( \partial S / \partial B \) exhibits a peak, $S$ in the high-field SC state above \( B^\star \) is higher compared to the low-field SC state.

 As shown by Yip et al. \cite{Yip1991}, the continuity of the free energy and its derivatives imposes strict constraints on phase diagrams: three second-order phase transition lines cannot meet at a single point. Since the SC phase transition to normal state at $B_{\rm c2}$ is always of second-order, the transition line at $B^\star$ cannot be of second-order. In contrast, it is permissible for the two SC states to be separated by a first-order phase transition line.  In this case, a constraint is that the slope of $B_{\rm c2}(T)$ for the high-field state must be steeper than that for the low-field state \cite{Yip1991}. This is consistent with the observed \( B_{\rm c2}(T) \)  line.

We further explore the phenomena occurring at $B^\star$. We observe that the anomaly sharpens within the SC state, whereas anomalies associated with Fermi surface instabilities, such as the MM crossover, would typically weaken due to the opening of the SC gap. This observation indicates that the transition cannot be attributed to the persistence of the metamagnetic crossover within the SC state.  Instead, the slope change in $B_{\rm c2}(T)$ at $B^\star$ indicates that the superconducting state changes its nature. Together with the thermodynamic anomaly, it points to a transition between two distinct superconducting states. Given that the $B^\star$ anomaly appears at the same field as the metamagnetic crossover (Fig. 3(c)), our results strongly suggest that the metamagnetism induces the superconducting phase transition.

Our results show that the entropy increases at $B^\star$. Since entropy in the SC state originates from quasiparticle excitations, the higher entropy in the high-field SC state implies the presence of more quasiparticles. This, in turn, indicates a smaller gap minimum, and thus greater anisotropy in the superconducting gap of the high-field phase (Fig. 4(b)).

For \( B \parallel a \)-axis, the symmetry is reduced to \( C_{\rm 2h} \), which supports two odd-parity irreducible representations: \( A_{\rm u}^{\rm 2h} = A_{\rm u} + B_{\rm 3u} \) and \( B_{\rm u}^{\rm 2h} = B_{\rm 1u} + B_{\rm 2u} \) \cite{Ishizuka2019, aoki2022b}. Thus, the simplest picture might be that the phase transition occurs between \( A_{\rm u}^{\rm 2h} \) and \( B_{\rm u}^{\rm 2h} \). However, it seems unlikely that the weak metamagnetism shown in the inset of Fig. \ref{phase}(b) would induce a drastic change in symmetry. Such a transition would inevitably be a strong first-order phase transition, involving a complete change from one order parameter to another (See also the Supplementary Fig. S4 for an illustration of the strong first-order phase transition for this case \cite{SM}).  Below, we discuss how the key lies in the flexibility of multi-component superconductivity, which can theoretically adjust its order parameter to an infinitesimally small external perturbation.

Given its first-order nature, the phase transition can occur between SC states with the same symmetry. For generality, we label the two order parameter components under magnetic field along the $a$-axis as $OP$ and $OP'$. The respective weights of these components, defined as $|\Delta_{OP}|^2/(|\Delta_{OP}|^2+|\Delta_{OP'}|^2)$ and $|\Delta_{OP'}|^2/(|\Delta_{OP}|^2+|\Delta_{OP'}|^2)$, can shift abruptly in response to external perturbations such as metamagnetism (Fig. \ref{entropy}(b)). This leads to a phase transition between states of the same symmetry.

While several scenarios may be possible, the simplest explanation consistent with our observations is a phase transition between two $A_{\rm u}^{\rm 2h}$ states, where $OP$ and $OP'$ correspond to the $A_{\rm u}$ and $B_{\rm 3u}$ states, respectively. This is supported by the following reasons. (1) The observed slope change of \( B_{\rm c2}(T) \) at \( B^\star \) (Supplementary Fig. S2 \cite{SM}), along with the consequent anomalous enhancement of \( B_{\rm c2} \), is consistent with the increased weight of the \( B_{\rm 3u} \) component. Since the \( B_{\rm 3u} \) state possesses the spin component along the \( a \)-axis \cite{Fujibayashi2022},  increasing its weight allows the system to gain Zeeman energy, thereby stabilizing the high-field superconducting state.
(2) Because $B_{\rm 3u}$ is point-nodal, while $A_{\rm u}$ is fully gapped, the dominance of $B_{\rm 3u}$ component in the high-field SC state causes a stronger anisotropy in the SC gap, consistent with our results.

An alternative scenario is  a phase transition between the two \( B_u^{2h} \) states. If the Fermi surface is two-dimensional, the \( B_{1u} \) state is expected to be fully gapped. In this case, an increasing contribution from the \( B_{2u} \) state at higher fields would enhance the gap anisotropy, which is broadly consistent with our observations. On the other hand, if the Fermi surface is three-dimensional, the situation becomes more complex; it is not straightforward to determine whether the \( B_{1u} \) or the \( B_{2u} \) state would be more isotropic. Moreover, the gap anisotropy could be altered by a possible change in the Fermi surface associated with the metamagnetic transition. Fully distinguishing among all possible scenarios requires rigorous theoretical investigations or experimental probes sensitive to the positions of the gap minimum, which we leave to future studies.



In summary, using ultra-clean single crystals of UTe$_2$, we investigated the interplay between metamagnetism and superconductivity for the field along the $a$-axis. Intriguingly, the metamagnetic crossover in the normal-conducting state evolves into a sharp anomaly in the SC state, signaling a first-order phase transition below $T_c$. This separates the two SC states with a step-wise increase of entropy at the transition. The emergence of high-field SC state in ultra-high quality crystals causes two-fold increase of the upper critical field despite only a 25\% increase of $T_c(0)$ compared to low-$T_c$ samples. Our results are best explained by  a transition between two superconducting states, accompanied by enhanced gap anisotropy in the high-field  state,  where the self-reconstruction of the order parameters occurs to adapt multi-component superconducting states to external perturbations.

We thank M. Garst, and T. Takimoto for stimulating discussions. We thank M. Nagai and K. Shirasaki for their experimental support. The work was supported by JSPS KAKENHI Grant Numbers JP16KK0106, JP17K05522, JP17K05529, JP20K03852, JP24K00590, JP24KK0062, JP23H04871, JP23H01132, JP23K03332 and 23K25829 and by the JAEA REIMEI Research Program. This work (A part of high magnetic field experiments) was performed at HFLSM under the IMR-GIMRT program (Proposal Numbers 202012-HMKPB-0012, 202112-HMKPB-0010, 202112-RDKGE-0036 and 202012-RDKGE-0084).

The data that support the findings of this study are available within the article and its Supplemental Material \cite{SM}.

\newpage

\widetext
\clearpage
\begin{center}

\textbf{\large Supplementary Information: Self-reconstruction of order parameter in spin-triplet superconductor UTe$_2$}

\end{center}

\setcounter{equation}{0}
\setcounter{figure}{0}
\setcounter{table}{0}
\setcounter{page}{1}
\setcounter{section}{0}
\makeatletter
\renewcommand{\theequation}{S\arabic{equation}}
\renewcommand{\thefigure}{S\arabic{figure}}

\begin{figure}[h]
\includegraphics[width=\linewidth,keepaspectratio]{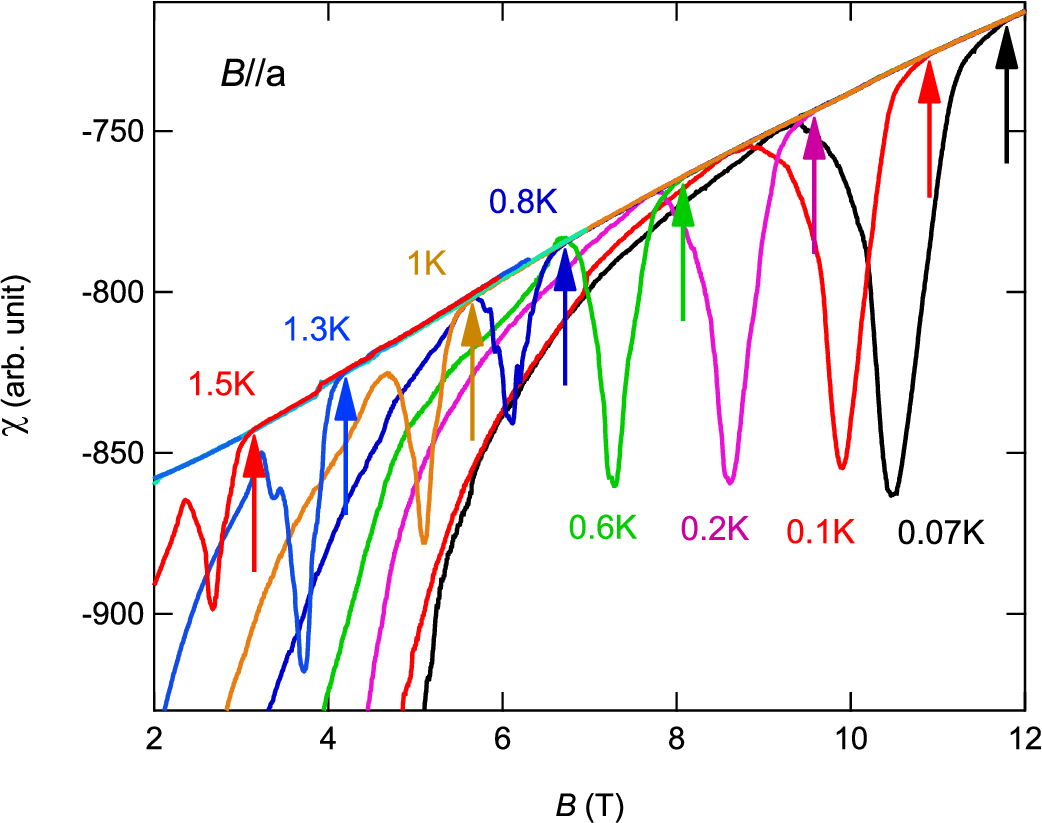}%
\caption{\label{chi}Determination of the upper critical field $B_{\rm c2}$ of superconductivity in UTe$_2$ for the field along $a$-axis by AC-magnetic susceptibility $\chi$. The arrows indicate $B_{\rm c2}$.}
\end{figure}

\newpage

\begin{figure}
\includegraphics[width=\linewidth,keepaspectratio]{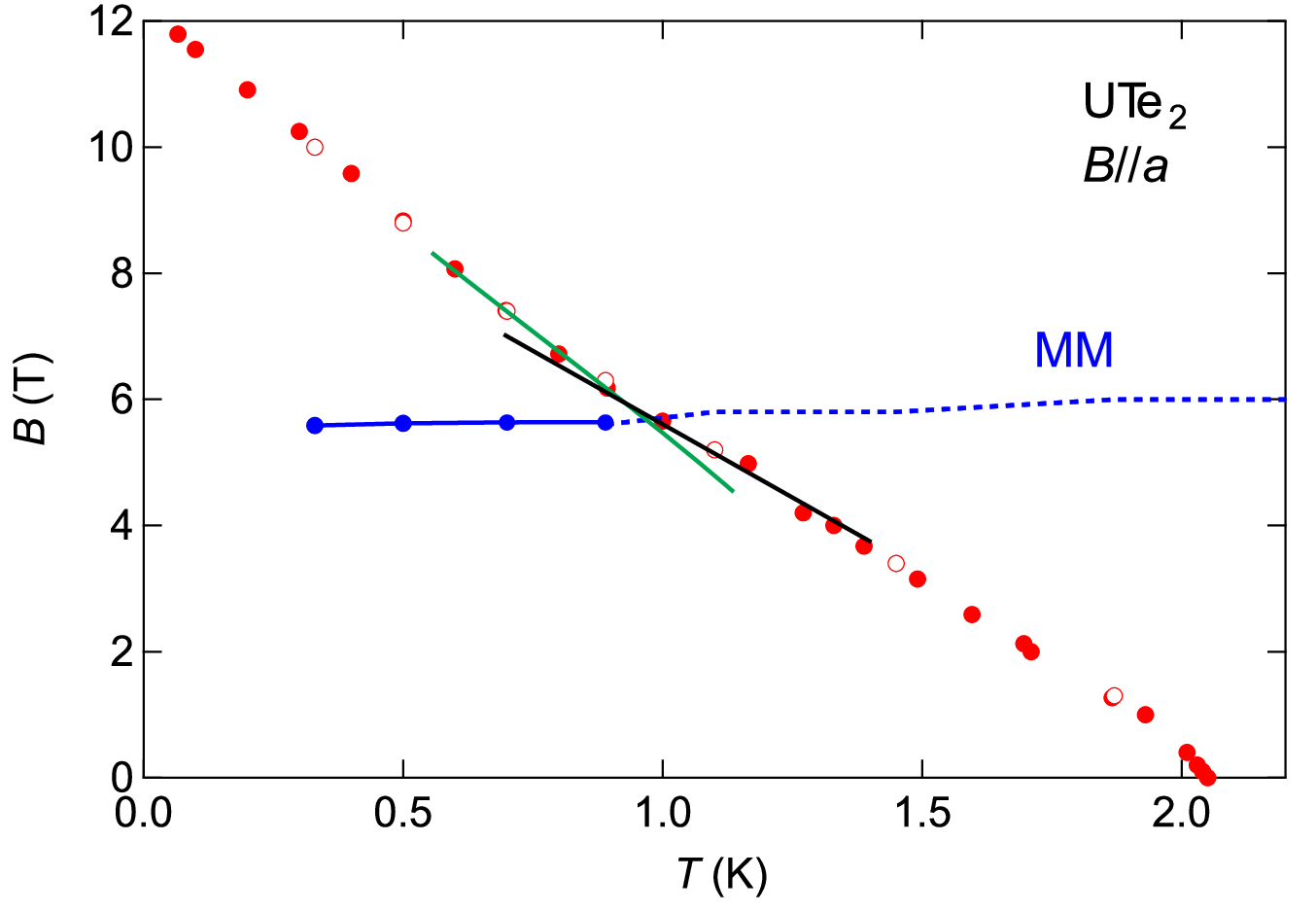}%
\caption{Superconducting phase diagram of UTe$_2$ for magnetic field applied along the $a$-axis. The data are identical to those shown in Fig. 1(b) of the main text. The two solid lines serve as guides to the eye.}
\end{figure}

\newpage

\begin{figure}
\includegraphics[width=\linewidth,keepaspectratio]{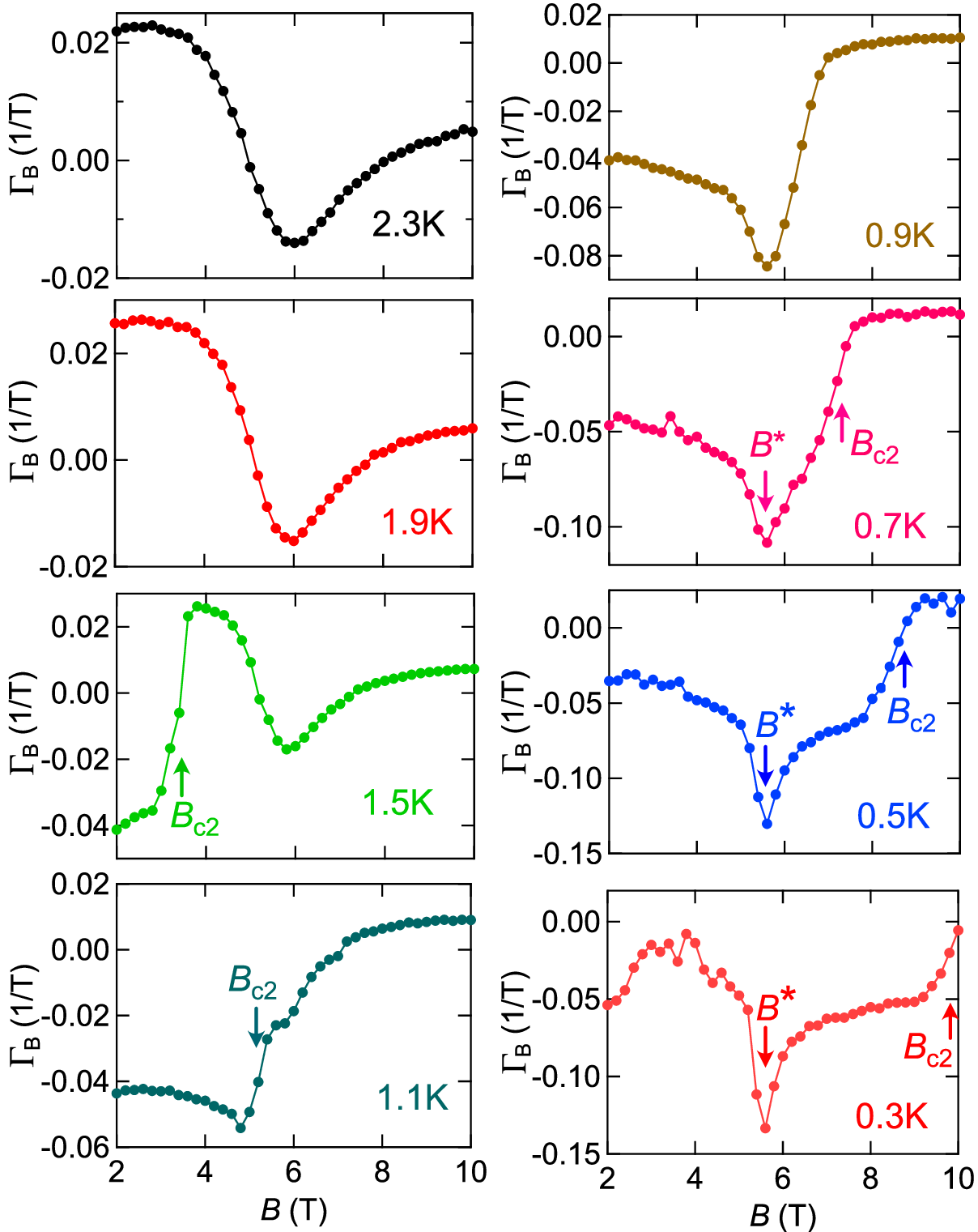}%
\caption{\label{MCE}Magnetic Gr\"{u}neisen parameter of UTe$_2$ as a function of magnetic field applied along the $a$-axis at various temperatures.}
\end{figure}

\newpage

\begin{figure}
\includegraphics[width=\linewidth,keepaspectratio]{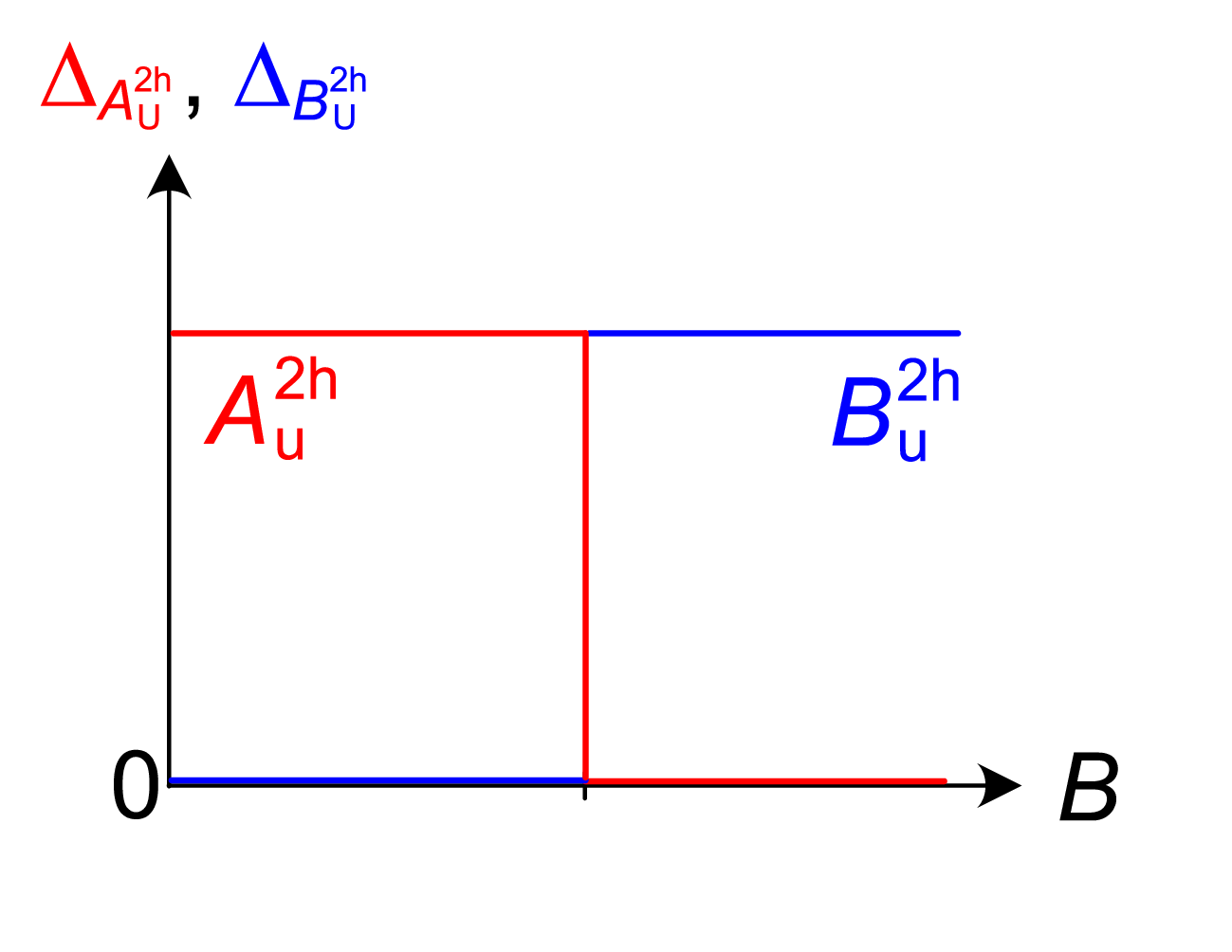}%
\caption{\label{GammaB_C}Schematic illustration of superconducting energy gap across a first-order phase transition between $A_u^{2h}$ and $B_u^{2h}$ states.}
\end{figure}


\begin{thebibliography}{44}%
\makeatletter
\providecommand \@ifxundefined [1]{%
 \@ifx{#1\undefined}
}%
\providecommand \@ifnum [1]{%
 \ifnum #1\expandafter \@firstoftwo
 \else \expandafter \@secondoftwo
 \fi
}%
\providecommand \@ifx [1]{%
 \ifx #1\expandafter \@firstoftwo
 \else \expandafter \@secondoftwo
 \fi
}%
\providecommand \natexlab [1]{#1}%
\providecommand \enquote  [1]{``#1''}%
\providecommand \bibnamefont  [1]{#1}%
\providecommand \bibfnamefont [1]{#1}%
\providecommand \citenamefont [1]{#1}%
\providecommand \href@noop [0]{\@secondoftwo}%
\providecommand \href [0]{\begingroup \@sanitize@url \@href}%
\providecommand \@href[1]{\@@startlink{#1}\@@href}%
\providecommand \@@href[1]{\endgroup#1\@@endlink}%
\providecommand \@sanitize@url [0]{\catcode `\\12\catcode `\$12\catcode
  `\&12\catcode `\#12\catcode `\^12\catcode `\_12\catcode `\%12\relax}%
\providecommand \@@startlink[1]{}%
\providecommand \@@endlink[0]{}%
\providecommand \url  [0]{\begingroup\@sanitize@url \@url }%
\providecommand \@url [1]{\endgroup\@href {#1}{\urlprefix }}%
\providecommand \urlprefix  [0]{URL }%
\providecommand \Eprint [0]{\href }%
\providecommand \doibase [0]{https://doi.org/}%
\providecommand \selectlanguage [0]{\@gobble}%
\providecommand \bibinfo  [0]{\@secondoftwo}%
\providecommand \bibfield  [0]{\@secondoftwo}%
\providecommand \translation [1]{[#1]}%
\providecommand \BibitemOpen [0]{}%
\providecommand \bibitemStop [0]{}%
\providecommand \bibitemNoStop [0]{.\EOS\space}%
\providecommand \EOS [0]{\spacefactor3000\relax}%
\providecommand \BibitemShut  [1]{\csname bibitem#1\endcsname}%
\let\auto@bib@innerbib\@empty
\bibitem [{\citenamefont {Joynt}\ and\ \citenamefont
  {Taillefer}(2002)}]{Joynt2002}%
  \BibitemOpen
  \bibfield  {author} {\bibinfo {author} {\bibfnamefont {R.}~\bibnamefont
  {Joynt}}\ and\ \bibinfo {author} {\bibfnamefont {L.}~\bibnamefont
  {Taillefer}},\ }\bibfield  {title} {\bibinfo {title} {The superconducting
  phases of {UPt}$_3$},\ }\href {https://doi.org/10.1103/RevModPhys.74.235}
  {\bibfield  {journal} {\bibinfo  {journal} {Rev. Mod. Phys.}\ }\textbf
  {\bibinfo {volume} {74}},\ \bibinfo {pages} {235} (\bibinfo {year}
  {2002})}\BibitemShut {NoStop}%
\bibitem [{\citenamefont {Khim}\ \emph {et~al.}(2021)\citenamefont {Khim},
  \citenamefont {Landaeta}, \citenamefont {Banda}, \citenamefont {Bannor},
  \citenamefont {Brando}, \citenamefont {Brydon}, \citenamefont {Hafner},
  \citenamefont {Küchler}, \citenamefont {Cardoso-Gil}, \citenamefont
  {Stockert}, \citenamefont {Mackenzie}, \citenamefont {Agterberg},
  \citenamefont {Geibel},\ and\ \citenamefont {Hassinger}}]{Khim2021}%
  \BibitemOpen
  \bibfield  {author} {\bibinfo {author} {\bibfnamefont {S.}~\bibnamefont
  {Khim}}, \bibinfo {author} {\bibfnamefont {J.~F.}\ \bibnamefont {Landaeta}},
  \bibinfo {author} {\bibfnamefont {J.}~\bibnamefont {Banda}}, \bibinfo
  {author} {\bibfnamefont {N.}~\bibnamefont {Bannor}}, \bibinfo {author}
  {\bibfnamefont {M.}~\bibnamefont {Brando}}, \bibinfo {author} {\bibfnamefont
  {P.~M.~R.}\ \bibnamefont {Brydon}}, \bibinfo {author} {\bibfnamefont
  {D.}~\bibnamefont {Hafner}}, \bibinfo {author} {\bibfnamefont
  {R.}~\bibnamefont {Küchler}}, \bibinfo {author} {\bibfnamefont
  {R.}~\bibnamefont {Cardoso-Gil}}, \bibinfo {author} {\bibfnamefont
  {U.}~\bibnamefont {Stockert}}, \bibinfo {author} {\bibfnamefont {A.~P.}\
  \bibnamefont {Mackenzie}}, \bibinfo {author} {\bibfnamefont {D.~F.}\
  \bibnamefont {Agterberg}}, \bibinfo {author} {\bibfnamefont {C.}~\bibnamefont
  {Geibel}},\ and\ \bibinfo {author} {\bibfnamefont {E.}~\bibnamefont
  {Hassinger}},\ }\bibfield  {title} {\bibinfo {title} {Field-induced
  transition within the superconducting state of {C}e{R}h$_2${A}s$_2$},\ }\href
  {https://doi.org/10.1126/science.abe7518} {\bibfield  {journal} {\bibinfo
  {journal} {Science}\ }\textbf {\bibinfo {volume} {373}},\ \bibinfo {pages}
  {1012} (\bibinfo {year} {2021})},\ \Eprint
  {https://arxiv.org/abs/https://www.science.org/doi/pdf/10.1126/science.abe7518}
  {https://www.science.org/doi/pdf/10.1126/science.abe7518} \BibitemShut
  {NoStop}%
\bibitem [{\citenamefont {Ran}\ \emph {et~al.}(2019{\natexlab{a}})\citenamefont
  {Ran}, \citenamefont {Eckberg}, \citenamefont {Ding}, \citenamefont
  {Furukawa}, \citenamefont {Metz}, \citenamefont {Saha}, \citenamefont {Liu},
  \citenamefont {Zic}, \citenamefont {Kim}, \citenamefont {Paglione},\ and\
  \citenamefont {Butch}}]{Ran2019}%
  \BibitemOpen
  \bibfield  {author} {\bibinfo {author} {\bibfnamefont {S.}~\bibnamefont
  {Ran}}, \bibinfo {author} {\bibfnamefont {C.}~\bibnamefont {Eckberg}},
  \bibinfo {author} {\bibfnamefont {Q.-P.}\ \bibnamefont {Ding}}, \bibinfo
  {author} {\bibfnamefont {Y.}~\bibnamefont {Furukawa}}, \bibinfo {author}
  {\bibfnamefont {T.}~\bibnamefont {Metz}}, \bibinfo {author} {\bibfnamefont
  {S.~R.}\ \bibnamefont {Saha}}, \bibinfo {author} {\bibfnamefont {I.-L.}\
  \bibnamefont {Liu}}, \bibinfo {author} {\bibfnamefont {M.}~\bibnamefont
  {Zic}}, \bibinfo {author} {\bibfnamefont {H.}~\bibnamefont {Kim}}, \bibinfo
  {author} {\bibfnamefont {J.}~\bibnamefont {Paglione}},\ and\ \bibinfo
  {author} {\bibfnamefont {N.~P.}\ \bibnamefont {Butch}},\ }\bibfield  {title}
  {\bibinfo {title} {Nearly ferromagnetic spin-triplet superconductivity},\
  }\href {https://doi.org/10.1126/science.aav8645} {\bibfield  {journal}
  {\bibinfo  {journal} {Science}\ }\textbf {\bibinfo {volume} {365}},\ \bibinfo
  {pages} {684} (\bibinfo {year} {2019}{\natexlab{a}})}\BibitemShut {NoStop}%
\bibitem [{\citenamefont {Aoki}\ \emph {et~al.}(2020)\citenamefont {Aoki},
  \citenamefont {Honda}, \citenamefont {Knebel}, \citenamefont {Braithwaite},
  \citenamefont {Nakamura}, \citenamefont {Li}, \citenamefont {Homma},
  \citenamefont {Shimizu}, \citenamefont {Sato}, \citenamefont {Brison},\ and\
  \citenamefont {Flouquet}}]{Aoki2020a}%
  \BibitemOpen
  \bibfield  {author} {\bibinfo {author} {\bibfnamefont {D.}~\bibnamefont
  {Aoki}}, \bibinfo {author} {\bibfnamefont {F.}~\bibnamefont {Honda}},
  \bibinfo {author} {\bibfnamefont {G.}~\bibnamefont {Knebel}}, \bibinfo
  {author} {\bibfnamefont {D.}~\bibnamefont {Braithwaite}}, \bibinfo {author}
  {\bibfnamefont {A.}~\bibnamefont {Nakamura}}, \bibinfo {author}
  {\bibfnamefont {D.}~\bibnamefont {Li}}, \bibinfo {author} {\bibfnamefont
  {Y.}~\bibnamefont {Homma}}, \bibinfo {author} {\bibfnamefont
  {Y.}~\bibnamefont {Shimizu}}, \bibinfo {author} {\bibfnamefont {Y.~J.}\
  \bibnamefont {Sato}}, \bibinfo {author} {\bibfnamefont {J.-P.}\ \bibnamefont
  {Brison}},\ and\ \bibinfo {author} {\bibfnamefont {J.}~\bibnamefont
  {Flouquet}},\ }\bibfield  {title} {\bibinfo {title} {Multiple superconducting
  phases and unusual enhancement of the upper critical field in {UTe}$_{2}$},\
  }\href {https://doi.org/10.7566/"J. Phys. Soc. Jpn.".89.053705} {\bibfield
  {journal} {\bibinfo  {journal} {Journal of the Physical Society of Japan}\
  }\textbf {\bibinfo {volume} {89}},\ \bibinfo {pages} {053705} (\bibinfo
  {year} {2020})}\BibitemShut {NoStop}%
\bibitem [{\citenamefont {Rosuel}\ \emph {et~al.}(2023)\citenamefont {Rosuel},
  \citenamefont {Marcenat}, \citenamefont {Knebel}, \citenamefont {Klein},
  \citenamefont {Pourret}, \citenamefont {Marquardt}, \citenamefont {Niu},
  \citenamefont {Rousseau}, \citenamefont {Demuer}, \citenamefont {Seyfarth},
  \citenamefont {Lapertot}, \citenamefont {Aoki}, \citenamefont {Braithwaite},
  \citenamefont {Flouquet},\ and\ \citenamefont {Brison}}]{Rosuel2022}%
  \BibitemOpen
  \bibfield  {author} {\bibinfo {author} {\bibfnamefont {A.}~\bibnamefont
  {Rosuel}}, \bibinfo {author} {\bibfnamefont {C.}~\bibnamefont {Marcenat}},
  \bibinfo {author} {\bibfnamefont {G.}~\bibnamefont {Knebel}}, \bibinfo
  {author} {\bibfnamefont {T.}~\bibnamefont {Klein}}, \bibinfo {author}
  {\bibfnamefont {A.}~\bibnamefont {Pourret}}, \bibinfo {author} {\bibfnamefont
  {N.}~\bibnamefont {Marquardt}}, \bibinfo {author} {\bibfnamefont
  {Q.}~\bibnamefont {Niu}}, \bibinfo {author} {\bibfnamefont {S.}~\bibnamefont
  {Rousseau}}, \bibinfo {author} {\bibfnamefont {A.}~\bibnamefont {Demuer}},
  \bibinfo {author} {\bibfnamefont {G.}~\bibnamefont {Seyfarth}}, \bibinfo
  {author} {\bibfnamefont {G.}~\bibnamefont {Lapertot}}, \bibinfo {author}
  {\bibfnamefont {D.}~\bibnamefont {Aoki}}, \bibinfo {author} {\bibfnamefont
  {D.}~\bibnamefont {Braithwaite}}, \bibinfo {author} {\bibfnamefont
  {J.}~\bibnamefont {Flouquet}},\ and\ \bibinfo {author} {\bibfnamefont
  {J.~P.}\ \bibnamefont {Brison}},\ }\bibfield  {title} {\bibinfo {title}
  {Field-induced tuning of the pairing state in a superconductor},\ }\href
  {https://doi.org/10.1103/PhysRevX.13.011022} {\bibfield  {journal} {\bibinfo
  {journal} {Phys. Rev. X}\ }\textbf {\bibinfo {volume} {13}},\ \bibinfo
  {pages} {011022} (\bibinfo {year} {2023})}\BibitemShut {NoStop}%
\bibitem [{\citenamefont {Sakai}\ \emph {et~al.}(2023)\citenamefont {Sakai},
  \citenamefont {Tokiwa}, \citenamefont {Opletal}, \citenamefont {Kimata},
  \citenamefont {Awaji}, \citenamefont {Sasaki}, \citenamefont {Aoki},
  \citenamefont {Kambe}, \citenamefont {Tokunaga},\ and\ \citenamefont
  {Haga}}]{Sakai2023}%
  \BibitemOpen
  \bibfield  {author} {\bibinfo {author} {\bibfnamefont {H.}~\bibnamefont
  {Sakai}}, \bibinfo {author} {\bibfnamefont {Y.}~\bibnamefont {Tokiwa}},
  \bibinfo {author} {\bibfnamefont {P.}~\bibnamefont {Opletal}}, \bibinfo
  {author} {\bibfnamefont {M.}~\bibnamefont {Kimata}}, \bibinfo {author}
  {\bibfnamefont {S.}~\bibnamefont {Awaji}}, \bibinfo {author} {\bibfnamefont
  {T.}~\bibnamefont {Sasaki}}, \bibinfo {author} {\bibfnamefont
  {D.}~\bibnamefont {Aoki}}, \bibinfo {author} {\bibfnamefont {S.}~\bibnamefont
  {Kambe}}, \bibinfo {author} {\bibfnamefont {Y.}~\bibnamefont {Tokunaga}},\
  and\ \bibinfo {author} {\bibfnamefont {Y.}~\bibnamefont {Haga}},\ }\bibfield
  {title} {\bibinfo {title} {Field induced multiple superconducting phases in
  {UT}e$_2$ along hard magnetic axis},\ }\href
  {https://doi.org/10.1103/PhysRevLett.130.196002} {\bibfield  {journal}
  {\bibinfo  {journal} {Phys. Rev. Lett.}\ }\textbf {\bibinfo {volume} {130}},\
  \bibinfo {pages} {196002} (\bibinfo {year} {2023})}\BibitemShut {NoStop}%
\bibitem [{\citenamefont {Kinjo}\ \emph {et~al.}(2023)\citenamefont {Kinjo},
  \citenamefont {Fujibayashi}, \citenamefont {Kitagawa}, \citenamefont
  {Ishida}, \citenamefont {Tokunaga}, \citenamefont {Sakai}, \citenamefont
  {Kambe}, \citenamefont {Nakamura}, \citenamefont {Shimizu}, \citenamefont
  {Homma}, \citenamefont {Li}, \citenamefont {Honda}, \citenamefont {Aoki},
  \citenamefont {Hiraki}, \citenamefont {Kimata},\ and\ \citenamefont
  {Sasaki}}]{Kinjo2022}%
  \BibitemOpen
  \bibfield  {author} {\bibinfo {author} {\bibfnamefont {K.}~\bibnamefont
  {Kinjo}}, \bibinfo {author} {\bibfnamefont {H.}~\bibnamefont {Fujibayashi}},
  \bibinfo {author} {\bibfnamefont {S.}~\bibnamefont {Kitagawa}}, \bibinfo
  {author} {\bibfnamefont {K.}~\bibnamefont {Ishida}}, \bibinfo {author}
  {\bibfnamefont {Y.}~\bibnamefont {Tokunaga}}, \bibinfo {author}
  {\bibfnamefont {H.}~\bibnamefont {Sakai}}, \bibinfo {author} {\bibfnamefont
  {S.}~\bibnamefont {Kambe}}, \bibinfo {author} {\bibfnamefont
  {A.}~\bibnamefont {Nakamura}}, \bibinfo {author} {\bibfnamefont
  {Y.}~\bibnamefont {Shimizu}}, \bibinfo {author} {\bibfnamefont
  {Y.}~\bibnamefont {Homma}}, \bibinfo {author} {\bibfnamefont {D.~X.}\
  \bibnamefont {Li}}, \bibinfo {author} {\bibfnamefont {F.}~\bibnamefont
  {Honda}}, \bibinfo {author} {\bibfnamefont {D.}~\bibnamefont {Aoki}},
  \bibinfo {author} {\bibfnamefont {K.}~\bibnamefont {Hiraki}}, \bibinfo
  {author} {\bibfnamefont {M.}~\bibnamefont {Kimata}},\ and\ \bibinfo {author}
  {\bibfnamefont {T.}~\bibnamefont {Sasaki}},\ }\bibfield  {title} {\bibinfo
  {title} {Change of superconducting character in {UTe}$_{2}$ induced by
  magnetic field},\ }\href {https://doi.org/10.1103/PhysRevB.107.L060502}
  {\bibfield  {journal} {\bibinfo  {journal} {Phys. Rev. B}\ }\textbf {\bibinfo
  {volume} {107}},\ \bibinfo {pages} {L060502} (\bibinfo {year}
  {2023})}\BibitemShut {NoStop}%
\bibitem [{\citenamefont {Braithwaite}\ \emph {et~al.}(2019)\citenamefont
  {Braithwaite}, \citenamefont {Vali{\v s}ka}, \citenamefont {Knebel},
  \citenamefont {Lapertot}, \citenamefont {Brison}, \citenamefont {Pourret},
  \citenamefont {Zhitomirsky}, \citenamefont {Flouquet}, \citenamefont
  {Honda},\ and\ \citenamefont {Aoki}}]{Braithwaite2019}%
  \BibitemOpen
  \bibfield  {author} {\bibinfo {author} {\bibfnamefont {D.}~\bibnamefont
  {Braithwaite}}, \bibinfo {author} {\bibfnamefont {M.}~\bibnamefont {Vali{\v
  s}ka}}, \bibinfo {author} {\bibfnamefont {G.}~\bibnamefont {Knebel}},
  \bibinfo {author} {\bibfnamefont {G.}~\bibnamefont {Lapertot}}, \bibinfo
  {author} {\bibfnamefont {J.~P.}\ \bibnamefont {Brison}}, \bibinfo {author}
  {\bibfnamefont {A.}~\bibnamefont {Pourret}}, \bibinfo {author} {\bibfnamefont
  {M.~E.}\ \bibnamefont {Zhitomirsky}}, \bibinfo {author} {\bibfnamefont
  {J.}~\bibnamefont {Flouquet}}, \bibinfo {author} {\bibfnamefont
  {F.}~\bibnamefont {Honda}},\ and\ \bibinfo {author} {\bibfnamefont
  {D.}~\bibnamefont {Aoki}},\ }\bibfield  {title} {\bibinfo {title} {Multiple
  superconducting phases in a nearly ferromagnetic system},\ }\href
  {https://doi.org/10.1038/s42005-019-0248-z} {\bibfield  {journal} {\bibinfo
  {journal} {Communications Physics}\ }\textbf {\bibinfo {volume} {2}},\
  \bibinfo {pages} {147} (\bibinfo {year} {2019})}\BibitemShut {NoStop}%
\bibitem [{\citenamefont {Knafo}\ \emph {et~al.}(2019)\citenamefont {Knafo},
  \citenamefont {Vali{\v s}ka}, \citenamefont {Braithwaite}, \citenamefont
  {Lapertot}, \citenamefont {Knebel}, \citenamefont {Pourret}, \citenamefont
  {Brison}, \citenamefont {Flouquet},\ and\ \citenamefont {Aoki}}]{Knafo2019}%
  \BibitemOpen
  \bibfield  {author} {\bibinfo {author} {\bibfnamefont {W.}~\bibnamefont
  {Knafo}}, \bibinfo {author} {\bibfnamefont {M.}~\bibnamefont {Vali{\v s}ka}},
  \bibinfo {author} {\bibfnamefont {D.}~\bibnamefont {Braithwaite}}, \bibinfo
  {author} {\bibfnamefont {G.}~\bibnamefont {Lapertot}}, \bibinfo {author}
  {\bibfnamefont {G.}~\bibnamefont {Knebel}}, \bibinfo {author} {\bibfnamefont
  {A.}~\bibnamefont {Pourret}}, \bibinfo {author} {\bibfnamefont {J.-P.}\
  \bibnamefont {Brison}}, \bibinfo {author} {\bibfnamefont {J.}~\bibnamefont
  {Flouquet}},\ and\ \bibinfo {author} {\bibfnamefont {D.}~\bibnamefont
  {Aoki}},\ }\bibfield  {title} {\bibinfo {title} {Magnetic-field-induced
  phenomena in the paramagnetic superconductor {UTe}$_2$},\ }\href
  {https://doi.org/10.7566/"J. Phys. Soc. Jpn.".88.063705} {\bibfield
  {journal} {\bibinfo  {journal} {Journal of the Physical Society of Japan}\
  }\textbf {\bibinfo {volume} {88}},\ \bibinfo {pages} {063705} (\bibinfo
  {year} {2019})}\BibitemShut {NoStop}%
\bibitem [{\citenamefont {Knafo}\ \emph {et~al.}(2021)\citenamefont {Knafo},
  \citenamefont {Nardone}, \citenamefont {Vali{\v s}ka}, \citenamefont
  {Zitouni}, \citenamefont {Lapertot}, \citenamefont {Aoki}, \citenamefont
  {Knebel},\ and\ \citenamefont {Braithwaite}}]{Knafo2021}%
  \BibitemOpen
  \bibfield  {author} {\bibinfo {author} {\bibfnamefont {W.}~\bibnamefont
  {Knafo}}, \bibinfo {author} {\bibfnamefont {M.}~\bibnamefont {Nardone}},
  \bibinfo {author} {\bibfnamefont {M.}~\bibnamefont {Vali{\v s}ka}}, \bibinfo
  {author} {\bibfnamefont {A.}~\bibnamefont {Zitouni}}, \bibinfo {author}
  {\bibfnamefont {G.}~\bibnamefont {Lapertot}}, \bibinfo {author}
  {\bibfnamefont {D.}~\bibnamefont {Aoki}}, \bibinfo {author} {\bibfnamefont
  {G.}~\bibnamefont {Knebel}},\ and\ \bibinfo {author} {\bibfnamefont
  {D.}~\bibnamefont {Braithwaite}},\ }\bibfield  {title} {\bibinfo {title}
  {Comparison of two superconducting phases induced by a magnetic field in
  {UTe}$_2$},\ }\href {https://doi.org/10.1038/s42005-021-00545-z} {\bibfield
  {journal} {\bibinfo  {journal} {Communications Physics}\ }\textbf {\bibinfo
  {volume} {4}},\ \bibinfo {pages} {40} (\bibinfo {year} {2021})}\BibitemShut
  {NoStop}%
\bibitem [{\citenamefont {Ran}\ \emph {et~al.}(2019{\natexlab{b}})\citenamefont
  {Ran}, \citenamefont {Liu}, \citenamefont {Eo}, \citenamefont {Campbell},
  \citenamefont {Neves}, \citenamefont {Fuhrman}, \citenamefont {Saha},
  \citenamefont {Eckberg}, \citenamefont {Kim}, \citenamefont {Graf},
  \citenamefont {Balakirev}, \citenamefont {Singleton}, \citenamefont
  {Paglione},\ and\ \citenamefont {Butch}}]{Ran2019a}%
  \BibitemOpen
  \bibfield  {author} {\bibinfo {author} {\bibfnamefont {S.}~\bibnamefont
  {Ran}}, \bibinfo {author} {\bibfnamefont {I.-L.}\ \bibnamefont {Liu}},
  \bibinfo {author} {\bibfnamefont {Y.~S.}\ \bibnamefont {Eo}}, \bibinfo
  {author} {\bibfnamefont {D.~J.}\ \bibnamefont {Campbell}}, \bibinfo {author}
  {\bibfnamefont {P.~M.}\ \bibnamefont {Neves}}, \bibinfo {author}
  {\bibfnamefont {W.~T.}\ \bibnamefont {Fuhrman}}, \bibinfo {author}
  {\bibfnamefont {S.~R.}\ \bibnamefont {Saha}}, \bibinfo {author}
  {\bibfnamefont {C.}~\bibnamefont {Eckberg}}, \bibinfo {author} {\bibfnamefont
  {H.}~\bibnamefont {Kim}}, \bibinfo {author} {\bibfnamefont {D.}~\bibnamefont
  {Graf}}, \bibinfo {author} {\bibfnamefont {F.}~\bibnamefont {Balakirev}},
  \bibinfo {author} {\bibfnamefont {J.}~\bibnamefont {Singleton}}, \bibinfo
  {author} {\bibfnamefont {J.}~\bibnamefont {Paglione}},\ and\ \bibinfo
  {author} {\bibfnamefont {N.~P.}\ \bibnamefont {Butch}},\ }\bibfield  {title}
  {\bibinfo {title} {Extreme magnetic field-boosted superconductivity},\ }\href
  {https://doi.org/10.1038/s41567-019-0670-x} {\bibfield  {journal} {\bibinfo
  {journal} {Nature Physics}\ }\textbf {\bibinfo {volume} {15}},\ \bibinfo
  {pages} {1250} (\bibinfo {year} {2019}{\natexlab{b}})}\BibitemShut {NoStop}%
\bibitem [{\citenamefont {Knebel}\ \emph {et~al.}(2019)\citenamefont {Knebel},
  \citenamefont {Knafo}, \citenamefont {Pourret}, \citenamefont {Niu},
  \citenamefont {Vali{\v s}ka}, \citenamefont {Braithwaite}, \citenamefont
  {Lapertot}, \citenamefont {Nardone}, \citenamefont {Zitouni}, \citenamefont
  {Mishra}, \citenamefont {Sheikin}, \citenamefont {Seyfarth}, \citenamefont
  {Brison}, \citenamefont {Aoki},\ and\ \citenamefont {Flouquet}}]{Knebel2019}%
  \BibitemOpen
  \bibfield  {author} {\bibinfo {author} {\bibfnamefont {G.}~\bibnamefont
  {Knebel}}, \bibinfo {author} {\bibfnamefont {W.}~\bibnamefont {Knafo}},
  \bibinfo {author} {\bibfnamefont {A.}~\bibnamefont {Pourret}}, \bibinfo
  {author} {\bibfnamefont {Q.}~\bibnamefont {Niu}}, \bibinfo {author}
  {\bibfnamefont {M.}~\bibnamefont {Vali{\v s}ka}}, \bibinfo {author}
  {\bibfnamefont {D.}~\bibnamefont {Braithwaite}}, \bibinfo {author}
  {\bibfnamefont {G.}~\bibnamefont {Lapertot}}, \bibinfo {author}
  {\bibfnamefont {M.}~\bibnamefont {Nardone}}, \bibinfo {author} {\bibfnamefont
  {A.}~\bibnamefont {Zitouni}}, \bibinfo {author} {\bibfnamefont
  {S.}~\bibnamefont {Mishra}}, \bibinfo {author} {\bibfnamefont
  {I.}~\bibnamefont {Sheikin}}, \bibinfo {author} {\bibfnamefont
  {G.}~\bibnamefont {Seyfarth}}, \bibinfo {author} {\bibfnamefont {J.-P.}\
  \bibnamefont {Brison}}, \bibinfo {author} {\bibfnamefont {D.}~\bibnamefont
  {Aoki}},\ and\ \bibinfo {author} {\bibfnamefont {J.}~\bibnamefont
  {Flouquet}},\ }\bibfield  {title} {\bibinfo {title} {Field-reentrant
  superconductivity close to a metamagnetic transition in the heavy-fermion
  superconductor {UTe}$_{2}$},\ }\href {https://doi.org/10.7566/"J. Phys. Soc.
  Jpn.".88.063707} {\bibfield  {journal} {\bibinfo  {journal} {Journal of the
  Physical Society of Japan}\ }\textbf {\bibinfo {volume} {88}},\ \bibinfo
  {pages} {063707} (\bibinfo {year} {2019})}\BibitemShut {NoStop}%
\bibitem [{\citenamefont {Haga}\ \emph {et~al.}(2022)\citenamefont {Haga},
  \citenamefont {Opletal}, \citenamefont {Tokiwa}, \citenamefont {Yamamoto},
  \citenamefont {Tokunaga}, \citenamefont {Kambe},\ and\ \citenamefont
  {Sakai}}]{Haga2022}%
  \BibitemOpen
  \bibfield  {author} {\bibinfo {author} {\bibfnamefont {Y.}~\bibnamefont
  {Haga}}, \bibinfo {author} {\bibfnamefont {P.}~\bibnamefont {Opletal}},
  \bibinfo {author} {\bibfnamefont {Y.}~\bibnamefont {Tokiwa}}, \bibinfo
  {author} {\bibfnamefont {E.}~\bibnamefont {Yamamoto}}, \bibinfo {author}
  {\bibfnamefont {Y.}~\bibnamefont {Tokunaga}}, \bibinfo {author}
  {\bibfnamefont {S.}~\bibnamefont {Kambe}},\ and\ \bibinfo {author}
  {\bibfnamefont {H.}~\bibnamefont {Sakai}},\ }\bibfield  {title} {\bibinfo
  {title} {Effect of uranium deficiency on normal and superconducting
  properties in unconventional superconductor {UTe}$_{2}$},\ }\href
  {http://iopscience.iop.org/article/10.1088/1361-648X/ac5201} {\bibfield
  {journal} {\bibinfo  {journal} {Journal of Physics: Condensed Matter}\
  }\textbf {\bibinfo {volume} {34}},\ \bibinfo {pages} {175601} (\bibinfo
  {year} {2022})}\BibitemShut {NoStop}%
\bibitem [{\citenamefont {Rosa}\ \emph {et~al.}(2022)\citenamefont {Rosa},
  \citenamefont {Weiland}, \citenamefont {Fender}, \citenamefont {Scott},
  \citenamefont {Ronning}, \citenamefont {Thompson}, \citenamefont {Bauer},\
  and\ \citenamefont {Thomas}}]{Rosa2022a}%
  \BibitemOpen
  \bibfield  {author} {\bibinfo {author} {\bibfnamefont {P.~F.~S.}\
  \bibnamefont {Rosa}}, \bibinfo {author} {\bibfnamefont {A.}~\bibnamefont
  {Weiland}}, \bibinfo {author} {\bibfnamefont {S.~S.}\ \bibnamefont {Fender}},
  \bibinfo {author} {\bibfnamefont {B.~L.}\ \bibnamefont {Scott}}, \bibinfo
  {author} {\bibfnamefont {F.}~\bibnamefont {Ronning}}, \bibinfo {author}
  {\bibfnamefont {J.~D.}\ \bibnamefont {Thompson}}, \bibinfo {author}
  {\bibfnamefont {E.~D.}\ \bibnamefont {Bauer}},\ and\ \bibinfo {author}
  {\bibfnamefont {S.~M.}\ \bibnamefont {Thomas}},\ }\bibfield  {title}
  {\bibinfo {title} {Single thermodynamic transition at 2 {K} in
  superconducting {UT}e$_2$ single crystals},\ }\href
  {https://doi.org/10.1038/s43246-022-00254-2} {\bibfield  {journal} {\bibinfo
  {journal} {Communications Materials}\ }\textbf {\bibinfo {volume} {3}},\
  \bibinfo {pages} {33} (\bibinfo {year} {2022})}\BibitemShut {NoStop}%
\bibitem [{\citenamefont {Aoki}\ \emph
  {et~al.}(2022{\natexlab{a}})\citenamefont {Aoki}, \citenamefont {Brison},
  \citenamefont {Flouquet}, \citenamefont {Ishida}, \citenamefont {Knebel},
  \citenamefont {Tokunaga},\ and\ \citenamefont {Yanase}}]{Aoki2022a}%
  \BibitemOpen
  \bibfield  {author} {\bibinfo {author} {\bibfnamefont {D.}~\bibnamefont
  {Aoki}}, \bibinfo {author} {\bibfnamefont {J.-P.}\ \bibnamefont {Brison}},
  \bibinfo {author} {\bibfnamefont {J.}~\bibnamefont {Flouquet}}, \bibinfo
  {author} {\bibfnamefont {K.}~\bibnamefont {Ishida}}, \bibinfo {author}
  {\bibfnamefont {G.}~\bibnamefont {Knebel}}, \bibinfo {author} {\bibfnamefont
  {Y.}~\bibnamefont {Tokunaga}},\ and\ \bibinfo {author} {\bibfnamefont
  {Y.}~\bibnamefont {Yanase}},\ }\bibfield  {title} {\bibinfo {title}
  {Unconventional superconductivity in {UTe}$_{2}$},\ }\href
  {http://iopscience.iop.org/article/10.1088/1361-648X/ac5863} {\bibfield
  {journal} {\bibinfo  {journal} {Journal of Physics: Condensed Matter}\
  }\textbf {\bibinfo {volume} {34}},\ \bibinfo {pages} {243002} (\bibinfo
  {year} {2022}{\natexlab{a}})}\BibitemShut {NoStop}%
\bibitem [{\citenamefont {Sakai}\ \emph {et~al.}(2022)\citenamefont {Sakai},
  \citenamefont {Opletal}, \citenamefont {Tokiwa}, \citenamefont {Yamamoto},
  \citenamefont {Tokunaga}, \citenamefont {Kambe},\ and\ \citenamefont
  {Haga}}]{Sakai2022}%
  \BibitemOpen
  \bibfield  {author} {\bibinfo {author} {\bibfnamefont {H.}~\bibnamefont
  {Sakai}}, \bibinfo {author} {\bibfnamefont {P.}~\bibnamefont {Opletal}},
  \bibinfo {author} {\bibfnamefont {Y.}~\bibnamefont {Tokiwa}}, \bibinfo
  {author} {\bibfnamefont {E.}~\bibnamefont {Yamamoto}}, \bibinfo {author}
  {\bibfnamefont {Y.}~\bibnamefont {Tokunaga}}, \bibinfo {author}
  {\bibfnamefont {S.}~\bibnamefont {Kambe}},\ and\ \bibinfo {author}
  {\bibfnamefont {Y.}~\bibnamefont {Haga}},\ }\bibfield  {title} {\bibinfo
  {title} {Single crystal growth of superconducting {UTe}$_{2}$ by molten salt
  flux method},\ }\href {https://doi.org/10.1103/PhysRevMaterials.6.073401}
  {\bibfield  {journal} {\bibinfo  {journal} {Phys. Rev. Materials}\ }\textbf
  {\bibinfo {volume} {6}},\ \bibinfo {pages} {073401} (\bibinfo {year}
  {2022})}\BibitemShut {NoStop}%
\bibitem [{\citenamefont {Aoki}\ \emph
  {et~al.}(2022{\natexlab{b}})\citenamefont {Aoki}, \citenamefont {Sakai},
  \citenamefont {Opletal}, \citenamefont {Tokiwa}, \citenamefont {Ishizuka},
  \citenamefont {Yanase}, \citenamefont {Harima}, \citenamefont {Nakamura},
  \citenamefont {Li}, \citenamefont {Homma}, \citenamefont {Shimizu},
  \citenamefont {Knebel}, \citenamefont {Flouquet},\ and\ \citenamefont
  {Haga}}]{Aoki2022c}%
  \BibitemOpen
  \bibfield  {author} {\bibinfo {author} {\bibfnamefont {D.}~\bibnamefont
  {Aoki}}, \bibinfo {author} {\bibfnamefont {H.}~\bibnamefont {Sakai}},
  \bibinfo {author} {\bibfnamefont {P.}~\bibnamefont {Opletal}}, \bibinfo
  {author} {\bibfnamefont {Y.}~\bibnamefont {Tokiwa}}, \bibinfo {author}
  {\bibfnamefont {J.}~\bibnamefont {Ishizuka}}, \bibinfo {author}
  {\bibfnamefont {Y.}~\bibnamefont {Yanase}}, \bibinfo {author} {\bibfnamefont
  {H.}~\bibnamefont {Harima}}, \bibinfo {author} {\bibfnamefont
  {A.}~\bibnamefont {Nakamura}}, \bibinfo {author} {\bibfnamefont
  {D.}~\bibnamefont {Li}}, \bibinfo {author} {\bibfnamefont {Y.}~\bibnamefont
  {Homma}}, \bibinfo {author} {\bibfnamefont {Y.}~\bibnamefont {Shimizu}},
  \bibinfo {author} {\bibfnamefont {G.}~\bibnamefont {Knebel}}, \bibinfo
  {author} {\bibfnamefont {J.}~\bibnamefont {Flouquet}},\ and\ \bibinfo
  {author} {\bibfnamefont {Y.}~\bibnamefont {Haga}},\ }\bibfield  {title}
  {\bibinfo {title} {First observation of the de haas–van alphen effect and
  fermi surfaces in the unconventional superconductor {UT}e$_2$},\ }\href
  {https://doi.org/10.7566/JPSJ.91.083704} {\bibfield  {journal} {\bibinfo
  {journal} {Journal of the Physical Society of Japan}\ }\textbf {\bibinfo
  {volume} {91}},\ \bibinfo {pages} {083704} (\bibinfo {year}
  {2022}{\natexlab{b}})}\BibitemShut {NoStop}%
\bibitem [{\citenamefont {Jiao}\ \emph {et~al.}(2020)\citenamefont {Jiao},
  \citenamefont {Howard}, \citenamefont {Ran}, \citenamefont {Wang},
  \citenamefont {Rodriguez}, \citenamefont {Sigrist}, \citenamefont {Wang},
  \citenamefont {Butch},\ and\ \citenamefont {Madhavan}}]{Jiao2020}%
  \BibitemOpen
  \bibfield  {author} {\bibinfo {author} {\bibfnamefont {L.}~\bibnamefont
  {Jiao}}, \bibinfo {author} {\bibfnamefont {S.}~\bibnamefont {Howard}},
  \bibinfo {author} {\bibfnamefont {S.}~\bibnamefont {Ran}}, \bibinfo {author}
  {\bibfnamefont {Z.}~\bibnamefont {Wang}}, \bibinfo {author} {\bibfnamefont
  {J.~O.}\ \bibnamefont {Rodriguez}}, \bibinfo {author} {\bibfnamefont
  {M.}~\bibnamefont {Sigrist}}, \bibinfo {author} {\bibfnamefont
  {Z.}~\bibnamefont {Wang}}, \bibinfo {author} {\bibfnamefont {N.~P.}\
  \bibnamefont {Butch}},\ and\ \bibinfo {author} {\bibfnamefont
  {V.}~\bibnamefont {Madhavan}},\ }\bibfield  {title} {\bibinfo {title} {Chiral
  superconductivity in heavy-fermion metal {UTe}$_2$},\ }\href
  {https://doi.org/10.1038/s41586-020-2122-2} {\bibfield  {journal} {\bibinfo
  {journal} {Nature}\ }\textbf {\bibinfo {volume} {579}},\ \bibinfo {pages}
  {523} (\bibinfo {year} {2020})}\BibitemShut {NoStop}%
\bibitem [{\citenamefont {Kittaka}\ \emph {et~al.}(2020)\citenamefont
  {Kittaka}, \citenamefont {Shimizu}, \citenamefont {Sakakibara}, \citenamefont
  {Nakamura}, \citenamefont {Li}, \citenamefont {Homma}, \citenamefont {Honda},
  \citenamefont {Aoki},\ and\ \citenamefont {Machida}}]{Kittaka2020}%
  \BibitemOpen
  \bibfield  {author} {\bibinfo {author} {\bibfnamefont {S.}~\bibnamefont
  {Kittaka}}, \bibinfo {author} {\bibfnamefont {Y.}~\bibnamefont {Shimizu}},
  \bibinfo {author} {\bibfnamefont {T.}~\bibnamefont {Sakakibara}}, \bibinfo
  {author} {\bibfnamefont {A.}~\bibnamefont {Nakamura}}, \bibinfo {author}
  {\bibfnamefont {D.}~\bibnamefont {Li}}, \bibinfo {author} {\bibfnamefont
  {Y.}~\bibnamefont {Homma}}, \bibinfo {author} {\bibfnamefont
  {F.}~\bibnamefont {Honda}}, \bibinfo {author} {\bibfnamefont
  {D.}~\bibnamefont {Aoki}},\ and\ \bibinfo {author} {\bibfnamefont
  {K.}~\bibnamefont {Machida}},\ }\bibfield  {title} {\bibinfo {title}
  {Orientation of point nodes and nonunitary triplet pairing tuned by the
  easy-axis magnetization in {UTe}$_2$},\ }\href
  {https://doi.org/10.1103/PhysRevResearch.2.032014} {\bibfield  {journal}
  {\bibinfo  {journal} {Phys. Rev. Research}\ }\textbf {\bibinfo {volume}
  {2}},\ \bibinfo {pages} {032014} (\bibinfo {year} {2020})}\BibitemShut
  {NoStop}%
\bibitem [{\citenamefont {Xu}\ \emph {et~al.}(2019)\citenamefont {Xu},
  \citenamefont {Sheng},\ and\ \citenamefont {Yang}}]{Xu2019}%
  \BibitemOpen
  \bibfield  {author} {\bibinfo {author} {\bibfnamefont {Y.}~\bibnamefont
  {Xu}}, \bibinfo {author} {\bibfnamefont {Y.}~\bibnamefont {Sheng}},\ and\
  \bibinfo {author} {\bibfnamefont {Y.-f.}\ \bibnamefont {Yang}},\ }\bibfield
  {title} {\bibinfo {title} {Quasi-two-dimensional fermi surfaces and unitary
  spin-triplet pairing in the heavy fermion superconductor {UT}e$_2$},\ }\href
  {https://doi.org/10.1103/PhysRevLett.123.217002} {\bibfield  {journal}
  {\bibinfo  {journal} {Phys. Rev. Lett.}\ }\textbf {\bibinfo {volume} {123}},\
  \bibinfo {pages} {217002} (\bibinfo {year} {2019})}\BibitemShut {NoStop}%
\bibitem [{\citenamefont {Ishizuka}\ \emph {et~al.}(2019)\citenamefont
  {Ishizuka}, \citenamefont {Sumita}, \citenamefont {Daido},\ and\
  \citenamefont {Yanase}}]{Ishizuka2019}%
  \BibitemOpen
  \bibfield  {author} {\bibinfo {author} {\bibfnamefont {J.}~\bibnamefont
  {Ishizuka}}, \bibinfo {author} {\bibfnamefont {S.}~\bibnamefont {Sumita}},
  \bibinfo {author} {\bibfnamefont {A.}~\bibnamefont {Daido}},\ and\ \bibinfo
  {author} {\bibfnamefont {Y.}~\bibnamefont {Yanase}},\ }\bibfield  {title}
  {\bibinfo {title} {Insulator-metal transition and topological
  superconductivity in {UT}e$_2$ from a first-principles calculation},\ }\href
  {https://doi.org/10.1103/PhysRevLett.123.217001} {\bibfield  {journal}
  {\bibinfo  {journal} {Phys. Rev. Lett.}\ }\textbf {\bibinfo {volume} {123}},\
  \bibinfo {pages} {217001} (\bibinfo {year} {2019})}\BibitemShut {NoStop}%
\bibitem [{\citenamefont {Machida}(2021)}]{Machida2021}%
  \BibitemOpen
  \bibfield  {author} {\bibinfo {author} {\bibfnamefont {K.}~\bibnamefont
  {Machida}},\ }\bibfield  {title} {\bibinfo {title} {Nonunitary triplet
  superconductivity tuned by field-controlled magnetization: {URhGe}, {UCoGe},
  and {UTe}$_{2}$},\ }\href {https://doi.org/10.1103/PhysRevB.104.014514}
  {\bibfield  {journal} {\bibinfo  {journal} {Phys. Rev. B}\ }\textbf {\bibinfo
  {volume} {104}},\ \bibinfo {pages} {014514} (\bibinfo {year}
  {2021})}\BibitemShut {NoStop}%
\bibitem [{\citenamefont {Ishihara}\ \emph {et~al.}(2023)\citenamefont
  {Ishihara}, \citenamefont {Roppongi}, \citenamefont {Kobayashi},
  \citenamefont {Imamura}, \citenamefont {Mizukami}, \citenamefont {Sakai},
  \citenamefont {Opletal}, \citenamefont {Tokiwa}, \citenamefont {Haga},
  \citenamefont {Hashimoto},\ and\ \citenamefont {Shibauchi}}]{Ishihara2023}%
  \BibitemOpen
  \bibfield  {author} {\bibinfo {author} {\bibfnamefont {K.}~\bibnamefont
  {Ishihara}}, \bibinfo {author} {\bibfnamefont {M.}~\bibnamefont {Roppongi}},
  \bibinfo {author} {\bibfnamefont {M.}~\bibnamefont {Kobayashi}}, \bibinfo
  {author} {\bibfnamefont {K.}~\bibnamefont {Imamura}}, \bibinfo {author}
  {\bibfnamefont {Y.}~\bibnamefont {Mizukami}}, \bibinfo {author}
  {\bibfnamefont {H.}~\bibnamefont {Sakai}}, \bibinfo {author} {\bibfnamefont
  {P.}~\bibnamefont {Opletal}}, \bibinfo {author} {\bibfnamefont
  {Y.}~\bibnamefont {Tokiwa}}, \bibinfo {author} {\bibfnamefont
  {Y.}~\bibnamefont {Haga}}, \bibinfo {author} {\bibfnamefont {K.}~\bibnamefont
  {Hashimoto}},\ and\ \bibinfo {author} {\bibfnamefont {T.}~\bibnamefont
  {Shibauchi}},\ }\bibfield  {title} {\bibinfo {title} {Chiral
  superconductivity in {UT}e$_2$ probed by anisotropic low-energy
  excitations},\ }\href {https://doi.org/10.1038/s41467-023-38688-y} {\bibfield
   {journal} {\bibinfo  {journal} {Nature Communications}\ }\textbf {\bibinfo
  {volume} {14}},\ \bibinfo {pages} {2966} (\bibinfo {year}
  {2023})}\BibitemShut {NoStop}%
\bibitem [{\citenamefont {Kanasugi}\ and\ \citenamefont
  {Yanase}(2022)}]{Kanasugi2022}%
  \BibitemOpen
  \bibfield  {author} {\bibinfo {author} {\bibfnamefont {S.}~\bibnamefont
  {Kanasugi}}\ and\ \bibinfo {author} {\bibfnamefont {Y.}~\bibnamefont
  {Yanase}},\ }\bibfield  {title} {\bibinfo {title} {Anapole superconductivity
  from pt-symmetric mixed-parity interband pairing},\ }\href
  {https://doi.org/10.1038/s42005-022-00804-7} {\bibfield  {journal} {\bibinfo
  {journal} {Communications Physics}\ }\textbf {\bibinfo {volume} {5}},\
  \bibinfo {pages} {39} (\bibinfo {year} {2022})}\BibitemShut {NoStop}%
\bibitem [{\citenamefont {Lee}\ \emph {et~al.}(2023)\citenamefont {Lee},
  \citenamefont {Woods}, \citenamefont {Rosa}, \citenamefont {Thomas},
  \citenamefont {Bauer}, \citenamefont {Lin},\ and\ \citenamefont
  {Movshovich}}]{Lee2023_arxiv}%
  \BibitemOpen
  \bibfield  {author} {\bibinfo {author} {\bibfnamefont {S.}~\bibnamefont
  {Lee}}, \bibinfo {author} {\bibfnamefont {A.~J.}\ \bibnamefont {Woods}},
  \bibinfo {author} {\bibfnamefont {P.~F.~S.}\ \bibnamefont {Rosa}}, \bibinfo
  {author} {\bibfnamefont {S.~M.}\ \bibnamefont {Thomas}}, \bibinfo {author}
  {\bibfnamefont {E.~D.}\ \bibnamefont {Bauer}}, \bibinfo {author}
  {\bibfnamefont {S.-Z.}\ \bibnamefont {Lin}},\ and\ \bibinfo {author}
  {\bibfnamefont {R.}~\bibnamefont {Movshovich}},\ }\bibfield  {title}
  {\bibinfo {title} {Anisotropic field-induced changes in the superconducting
  order parameter of {UT}e$_2$. {P}reprint at
  https://doi.org/10.48550/arxiv.2310.04938},\ }\href
  {https://doi.org/10.48550/arXiv.2310.04938} {\  (\bibinfo {year}
  {2023})}\BibitemShut {NoStop}%
\bibitem [{\citenamefont {Ajeesh}\ \emph {et~al.}(2023)\citenamefont {Ajeesh},
  \citenamefont {Bordelon}, \citenamefont {Girod}, \citenamefont {Mishra},
  \citenamefont {Ronning}, \citenamefont {Bauer}, \citenamefont {Maiorov},
  \citenamefont {Thompson}, \citenamefont {Rosa},\ and\ \citenamefont
  {Thomas}}]{Ajeesh2023}%
  \BibitemOpen
  \bibfield  {author} {\bibinfo {author} {\bibfnamefont {M.~O.}\ \bibnamefont
  {Ajeesh}}, \bibinfo {author} {\bibfnamefont {M.}~\bibnamefont {Bordelon}},
  \bibinfo {author} {\bibfnamefont {C.}~\bibnamefont {Girod}}, \bibinfo
  {author} {\bibfnamefont {S.}~\bibnamefont {Mishra}}, \bibinfo {author}
  {\bibfnamefont {F.}~\bibnamefont {Ronning}}, \bibinfo {author} {\bibfnamefont
  {E.~D.}\ \bibnamefont {Bauer}}, \bibinfo {author} {\bibfnamefont
  {B.}~\bibnamefont {Maiorov}}, \bibinfo {author} {\bibfnamefont {J.~D.}\
  \bibnamefont {Thompson}}, \bibinfo {author} {\bibfnamefont {P.~F.~S.}\
  \bibnamefont {Rosa}},\ and\ \bibinfo {author} {\bibfnamefont {S.~M.}\
  \bibnamefont {Thomas}},\ }\bibfield  {title} {\bibinfo {title} {Fate of
  time-reversal symmetry breaking in {UTe}$_{2}$},\ }\href
  {https://doi.org/10.1103/PhysRevX.13.041019} {\bibfield  {journal} {\bibinfo
  {journal} {Phys. Rev. X}\ }\textbf {\bibinfo {volume} {13}},\ \bibinfo
  {pages} {041019} (\bibinfo {year} {2023})}\BibitemShut {NoStop}%
\bibitem [{\citenamefont {Suetsugu}\ \emph {et~al.}(2024)\citenamefont
  {Suetsugu}, \citenamefont {Shimomura}, \citenamefont {Kamimura},
  \citenamefont {Asaba}, \citenamefont {Asaeda}, \citenamefont {Kosuge},
  \citenamefont {Sekino}, \citenamefont {Ikemori}, \citenamefont {Kasahara},
  \citenamefont {Kohsaka}, \citenamefont {Lee}, \citenamefont {Yanase},
  \citenamefont {Sakai}, \citenamefont {Opletal}, \citenamefont {Tokiwa},
  \citenamefont {Haga},\ and\ \citenamefont {Matsuda}}]{Suetsugu2023}%
  \BibitemOpen
  \bibfield  {author} {\bibinfo {author} {\bibfnamefont {S.}~\bibnamefont
  {Suetsugu}}, \bibinfo {author} {\bibfnamefont {M.}~\bibnamefont {Shimomura}},
  \bibinfo {author} {\bibfnamefont {M.}~\bibnamefont {Kamimura}}, \bibinfo
  {author} {\bibfnamefont {T.}~\bibnamefont {Asaba}}, \bibinfo {author}
  {\bibfnamefont {H.}~\bibnamefont {Asaeda}}, \bibinfo {author} {\bibfnamefont
  {Y.}~\bibnamefont {Kosuge}}, \bibinfo {author} {\bibfnamefont
  {Y.}~\bibnamefont {Sekino}}, \bibinfo {author} {\bibfnamefont
  {S.}~\bibnamefont {Ikemori}}, \bibinfo {author} {\bibfnamefont
  {Y.}~\bibnamefont {Kasahara}}, \bibinfo {author} {\bibfnamefont
  {Y.}~\bibnamefont {Kohsaka}}, \bibinfo {author} {\bibfnamefont
  {M.}~\bibnamefont {Lee}}, \bibinfo {author} {\bibfnamefont {Y.}~\bibnamefont
  {Yanase}}, \bibinfo {author} {\bibfnamefont {H.}~\bibnamefont {Sakai}},
  \bibinfo {author} {\bibfnamefont {P.}~\bibnamefont {Opletal}}, \bibinfo
  {author} {\bibfnamefont {Y.}~\bibnamefont {Tokiwa}}, \bibinfo {author}
  {\bibfnamefont {Y.}~\bibnamefont {Haga}},\ and\ \bibinfo {author}
  {\bibfnamefont {Y.}~\bibnamefont {Matsuda}},\ }\bibfield  {title} {\bibinfo
  {title} {Fully gapped pairing state in spin-triplet superconductor
  {UTe}$_2$},\ }\href {https://doi.org/10.1126/sciadv.adk3772} {\bibfield
  {journal} {\bibinfo  {journal} {Science Advances}\ }\textbf {\bibinfo
  {volume} {10}},\ \bibinfo {pages} {eadk3772} (\bibinfo {year} {2024})},\
  \Eprint
  {https://arxiv.org/abs/https://www.science.org/doi/pdf/10.1126/sciadv.adk3772}
  {https://www.science.org/doi/pdf/10.1126/sciadv.adk3772} \BibitemShut
  {NoStop}%
\bibitem [{\citenamefont {Matsumura}\ \emph {et~al.}(2023)\citenamefont
  {Matsumura}, \citenamefont {Fujibayashi}, \citenamefont {Kinjo},
  \citenamefont {Kitagawa}, \citenamefont {Ishida}, \citenamefont {Tokunaga},
  \citenamefont {Sakai}, \citenamefont {Kambe}, \citenamefont {Nakamura},
  \citenamefont {Shimizu}, \citenamefont {Homma}, \citenamefont {Li},
  \citenamefont {Honda},\ and\ \citenamefont {Aoki}}]{Matsumura2023}%
  \BibitemOpen
  \bibfield  {author} {\bibinfo {author} {\bibfnamefont {H.}~\bibnamefont
  {Matsumura}}, \bibinfo {author} {\bibfnamefont {H.}~\bibnamefont
  {Fujibayashi}}, \bibinfo {author} {\bibfnamefont {K.}~\bibnamefont {Kinjo}},
  \bibinfo {author} {\bibfnamefont {S.}~\bibnamefont {Kitagawa}}, \bibinfo
  {author} {\bibfnamefont {K.}~\bibnamefont {Ishida}}, \bibinfo {author}
  {\bibfnamefont {Y.}~\bibnamefont {Tokunaga}}, \bibinfo {author}
  {\bibfnamefont {H.}~\bibnamefont {Sakai}}, \bibinfo {author} {\bibfnamefont
  {S.}~\bibnamefont {Kambe}}, \bibinfo {author} {\bibfnamefont
  {A.}~\bibnamefont {Nakamura}}, \bibinfo {author} {\bibfnamefont
  {Y.}~\bibnamefont {Shimizu}}, \bibinfo {author} {\bibfnamefont
  {Y.}~\bibnamefont {Homma}}, \bibinfo {author} {\bibfnamefont
  {D.}~\bibnamefont {Li}}, \bibinfo {author} {\bibfnamefont {F.}~\bibnamefont
  {Honda}},\ and\ \bibinfo {author} {\bibfnamefont {D.}~\bibnamefont {Aoki}},\
  }\bibfield  {title} {\bibinfo {title} {Large reduction in the a-axis knight
  shift on {UT}e$_2$ with {T}$_c$ = 2.1 {K}},\ }\href
  {https://doi.org/10.7566/JPSJ.92.063701} {\bibfield  {journal} {\bibinfo
  {journal} {Journal of the Physical Society of Japan}\ }\textbf {\bibinfo
  {volume} {92}},\ \bibinfo {pages} {063701} (\bibinfo {year}
  {2023})}\BibitemShut {NoStop}%
\bibitem [{\citenamefont {Hayes}\ \emph {et~al.}(2024)\citenamefont {Hayes},
  \citenamefont {Metz}, \citenamefont {Frank}, \citenamefont {Saha},
  \citenamefont {Butch}, \citenamefont {Mishra}, \citenamefont {Hirschfeld},\
  and\ \citenamefont {Paglione}}]{Hayes2024}%
  \BibitemOpen
  \bibfield  {author} {\bibinfo {author} {\bibfnamefont {I.~M.}\ \bibnamefont
  {Hayes}}, \bibinfo {author} {\bibfnamefont {T.~E.}\ \bibnamefont {Metz}},
  \bibinfo {author} {\bibfnamefont {C.~E.}\ \bibnamefont {Frank}}, \bibinfo
  {author} {\bibfnamefont {S.~R.}\ \bibnamefont {Saha}}, \bibinfo {author}
  {\bibfnamefont {N.~P.}\ \bibnamefont {Butch}}, \bibinfo {author}
  {\bibfnamefont {V.}~\bibnamefont {Mishra}}, \bibinfo {author} {\bibfnamefont
  {P.~J.}\ \bibnamefont {Hirschfeld}},\ and\ \bibinfo {author} {\bibfnamefont
  {J.}~\bibnamefont {Paglione}},\ }\href {https://arxiv.org/abs/2402.19353}
  {\bibinfo {title} {Robust nodal behavior in the thermal conductivity of
  superconducting ute$_2$}} (\bibinfo {year} {2024}),\ \Eprint
  {https://arxiv.org/abs/2402.19353} {arXiv:2402.19353 [cond-mat.supr-con]}
  \BibitemShut {NoStop}%
\bibitem [{\citenamefont {Theuss}\ \emph {et~al.}(2024)\citenamefont {Theuss},
  \citenamefont {Shragai}, \citenamefont {Grissonnanche}, \citenamefont
  {Hayes}, \citenamefont {Saha}, \citenamefont {Eo}, \citenamefont {Suarez},
  \citenamefont {Shishidou}, \citenamefont {Butch}, \citenamefont {Paglione},\
  and\ \citenamefont {Ramshaw}}]{Theuss2024}%
  \BibitemOpen
  \bibfield  {author} {\bibinfo {author} {\bibfnamefont {F.}~\bibnamefont
  {Theuss}}, \bibinfo {author} {\bibfnamefont {A.}~\bibnamefont {Shragai}},
  \bibinfo {author} {\bibfnamefont {G.}~\bibnamefont {Grissonnanche}}, \bibinfo
  {author} {\bibfnamefont {I.~M.}\ \bibnamefont {Hayes}}, \bibinfo {author}
  {\bibfnamefont {S.~R.}\ \bibnamefont {Saha}}, \bibinfo {author}
  {\bibfnamefont {Y.~S.}\ \bibnamefont {Eo}}, \bibinfo {author} {\bibfnamefont
  {A.}~\bibnamefont {Suarez}}, \bibinfo {author} {\bibfnamefont
  {T.}~\bibnamefont {Shishidou}}, \bibinfo {author} {\bibfnamefont {N.~P.}\
  \bibnamefont {Butch}}, \bibinfo {author} {\bibfnamefont {J.}~\bibnamefont
  {Paglione}},\ and\ \bibinfo {author} {\bibfnamefont {B.~J.}\ \bibnamefont
  {Ramshaw}},\ }\bibfield  {title} {\bibinfo {title} {Single-component
  superconductivity in ute2 at ambient pressure},\ }\href
  {https://doi.org/10.1038/s41567-024-02493-1} {\bibfield  {journal} {\bibinfo
  {journal} {Nature Physics}\ }\textbf {\bibinfo {volume} {20}},\ \bibinfo
  {pages} {1124} (\bibinfo {year} {2024})}\BibitemShut {NoStop}%
\bibitem [{\citenamefont {Gu}\ \emph {et~al.}(2025)\citenamefont {Gu},
  \citenamefont {Wang}, \citenamefont {Carroll}, \citenamefont {Zhussupbekov},
  \citenamefont {Broyles}, \citenamefont {Ran}, \citenamefont {Butch},
  \citenamefont {Horn}, \citenamefont {Saha}, \citenamefont {Paglione},
  \citenamefont {Liu}, \citenamefont {Davis},\ and\ \citenamefont {Lee}}]{Gu}%
  \BibitemOpen
  \bibfield  {author} {\bibinfo {author} {\bibfnamefont {Q.}~\bibnamefont
  {Gu}}, \bibinfo {author} {\bibfnamefont {S.}~\bibnamefont {Wang}}, \bibinfo
  {author} {\bibfnamefont {J.~P.}\ \bibnamefont {Carroll}}, \bibinfo {author}
  {\bibfnamefont {K.}~\bibnamefont {Zhussupbekov}}, \bibinfo {author}
  {\bibfnamefont {C.}~\bibnamefont {Broyles}}, \bibinfo {author} {\bibfnamefont
  {S.}~\bibnamefont {Ran}}, \bibinfo {author} {\bibfnamefont {N.~P.}\
  \bibnamefont {Butch}}, \bibinfo {author} {\bibfnamefont {J.~A.}\ \bibnamefont
  {Horn}}, \bibinfo {author} {\bibfnamefont {S.}~\bibnamefont {Saha}}, \bibinfo
  {author} {\bibfnamefont {J.}~\bibnamefont {Paglione}}, \bibinfo {author}
  {\bibfnamefont {X.}~\bibnamefont {Liu}}, \bibinfo {author} {\bibfnamefont
  {J.~C.~S.}\ \bibnamefont {Davis}},\ and\ \bibinfo {author} {\bibfnamefont
  {D.-H.}\ \bibnamefont {Lee}},\ }\bibfield  {title} {\bibinfo {title} {Pair
  wave function symmetry in ute<sub>2</sub> from zero-energy surface state
  visualization},\ }\href {https://doi.org/10.1126/science.adk7219} {\bibfield
  {journal} {\bibinfo  {journal} {Science}\ }\textbf {\bibinfo {volume}
  {388}},\ \bibinfo {pages} {938} (\bibinfo {year} {2025})}\BibitemShut
  {NoStop}%
\bibitem [{\citenamefont {Tei}\ \emph {et~al.}(2024)\citenamefont {Tei},
  \citenamefont {Mizushima},\ and\ \citenamefont {Fujimoto}}]{Tei}%
  \BibitemOpen
  \bibfield  {author} {\bibinfo {author} {\bibfnamefont {J.}~\bibnamefont
  {Tei}}, \bibinfo {author} {\bibfnamefont {T.}~\bibnamefont {Mizushima}},\
  and\ \bibinfo {author} {\bibfnamefont {S.}~\bibnamefont {Fujimoto}},\
  }\bibfield  {title} {\bibinfo {title} {Pairing symmetries of multiple
  superconducting phases in ${\mathrm{ute}}_{2}$: Competition between
  ferromagnetic and antiferromagnetic fluctuations},\ }\href
  {https://doi.org/10.1103/PhysRevB.109.064516} {\bibfield  {journal} {\bibinfo
   {journal} {Phys. Rev. B}\ }\textbf {\bibinfo {volume} {109}},\ \bibinfo
  {pages} {064516} (\bibinfo {year} {2024})}\BibitemShut {NoStop}%
\bibitem [{\citenamefont {Niu}\ \emph {et~al.}(2020)\citenamefont {Niu},
  \citenamefont {Knebel}, \citenamefont {Braithwaite}, \citenamefont {Aoki},
  \citenamefont {Lapertot}, \citenamefont {Seyfarth}, \citenamefont {Brison},
  \citenamefont {Flouquet},\ and\ \citenamefont {Pourret}}]{Niu2020}%
  \BibitemOpen
  \bibfield  {author} {\bibinfo {author} {\bibfnamefont {Q.}~\bibnamefont
  {Niu}}, \bibinfo {author} {\bibfnamefont {G.}~\bibnamefont {Knebel}},
  \bibinfo {author} {\bibfnamefont {D.}~\bibnamefont {Braithwaite}}, \bibinfo
  {author} {\bibfnamefont {D.}~\bibnamefont {Aoki}}, \bibinfo {author}
  {\bibfnamefont {G.}~\bibnamefont {Lapertot}}, \bibinfo {author}
  {\bibfnamefont {G.}~\bibnamefont {Seyfarth}}, \bibinfo {author}
  {\bibfnamefont {J.-P.}\ \bibnamefont {Brison}}, \bibinfo {author}
  {\bibfnamefont {J.}~\bibnamefont {Flouquet}},\ and\ \bibinfo {author}
  {\bibfnamefont {A.}~\bibnamefont {Pourret}},\ }\bibfield  {title} {\bibinfo
  {title} {Fermi-surface instability in the heavy-fermion superconductor
  {UT}e$_2$},\ }\href {https://doi.org/10.1103/PhysRevLett.124.086601}
  {\bibfield  {journal} {\bibinfo  {journal} {Phys. Rev. Lett.}\ }\textbf
  {\bibinfo {volume} {124}},\ \bibinfo {pages} {086601} (\bibinfo {year}
  {2020})}\BibitemShut {NoStop}%
\bibitem [{\citenamefont {Tokiwa}\ \emph {et~al.}()\citenamefont {Tokiwa},
  \citenamefont {Opletal}, \citenamefont {Sakai}, \citenamefont {Yamamoto},
  \citenamefont {Kambe}, \citenamefont {Aoki}, \citenamefont {Tokunaga},\ and\
  \citenamefont {Haga}}]{SM}%
  \BibitemOpen
  \bibfield  {author} {\bibinfo {author} {\bibfnamefont {Y.}~\bibnamefont
  {Tokiwa}}, \bibinfo {author} {\bibfnamefont {P.}~\bibnamefont {Opletal}},
  \bibinfo {author} {\bibfnamefont {H.}~\bibnamefont {Sakai}}, \bibinfo
  {author} {\bibfnamefont {E.}~\bibnamefont {Yamamoto}}, \bibinfo {author}
  {\bibfnamefont {S.}~\bibnamefont {Kambe}}, \bibinfo {author} {\bibfnamefont
  {D.}~\bibnamefont {Aoki}}, \bibinfo {author} {\bibfnamefont {Y.}~\bibnamefont
  {Tokunaga}},\ and\ \bibinfo {author} {\bibfnamefont {Y.}~\bibnamefont
  {Haga}},\ }\href@noop {} {\bibinfo {title} {Suplementary
  material}}\BibitemShut {NoStop}%
\bibitem [{\citenamefont {Carbotte}(1990)}]{Carbotte}%
  \BibitemOpen
  \bibfield  {author} {\bibinfo {author} {\bibfnamefont {J.~P.}\ \bibnamefont
  {Carbotte}},\ }\bibfield  {title} {\bibinfo {title} {Properties of
  boson-exchange superconductors},\ }\href
  {https://doi.org/10.1103/RevModPhys.62.1027} {\bibfield  {journal} {\bibinfo
  {journal} {Rev. Mod. Phys.}\ }\textbf {\bibinfo {volume} {62}},\ \bibinfo
  {pages} {1027} (\bibinfo {year} {1990})}\BibitemShut {NoStop}%
\bibitem [{\citenamefont {Wu}\ \emph {et~al.}(2017)\citenamefont {Wu},
  \citenamefont {Bastien}, \citenamefont {Taupin}, \citenamefont {Paulsen},
  \citenamefont {Howald}, \citenamefont {Aoki},\ and\ \citenamefont
  {Brison}}]{Wu2017}%
  \BibitemOpen
  \bibfield  {author} {\bibinfo {author} {\bibfnamefont {B.}~\bibnamefont
  {Wu}}, \bibinfo {author} {\bibfnamefont {G.}~\bibnamefont {Bastien}},
  \bibinfo {author} {\bibfnamefont {M.}~\bibnamefont {Taupin}}, \bibinfo
  {author} {\bibfnamefont {C.}~\bibnamefont {Paulsen}}, \bibinfo {author}
  {\bibfnamefont {L.}~\bibnamefont {Howald}}, \bibinfo {author} {\bibfnamefont
  {D.}~\bibnamefont {Aoki}},\ and\ \bibinfo {author} {\bibfnamefont {J.-P.}\
  \bibnamefont {Brison}},\ }\bibfield  {title} {\bibinfo {title} {Pairing
  mechanism in the ferromagnetic superconductor ucoge},\ }\href
  {https://doi.org/10.1038/ncomms14480} {\bibfield  {journal} {\bibinfo
  {journal} {Nature Communications}\ }\textbf {\bibinfo {volume} {8}},\
  \bibinfo {pages} {14480} (\bibinfo {year} {2017})}\BibitemShut {NoStop}%
\bibitem [{\citenamefont {Tokiwa}\ and\ \citenamefont
  {Gegenwart}(2011)}]{Tokiwa2011}%
  \BibitemOpen
  \bibfield  {author} {\bibinfo {author} {\bibfnamefont {Y.}~\bibnamefont
  {Tokiwa}}\ and\ \bibinfo {author} {\bibfnamefont {P.}~\bibnamefont
  {Gegenwart}},\ }\bibfield  {title} {\bibinfo {title} {High-resolution
  alternating-field technique to determine the magnetocaloric effect of metals
  down to very low temperatures},\ }\href {https://doi.org/10.1063/1.3529433}
  {\bibfield  {journal} {\bibinfo  {journal} {Review of Scientific
  Instruments}\ }\textbf {\bibinfo {volume} {82}},\ \bibinfo {pages} {013905}
  (\bibinfo {year} {2011})}\BibitemShut {NoStop}%
\bibitem [{\citenamefont {Miyake}\ \emph {et~al.}(2019)\citenamefont {Miyake},
  \citenamefont {Shimizu}, \citenamefont {Sato}, \citenamefont {Li},
  \citenamefont {Nakamura}, \citenamefont {Homma}, \citenamefont {Honda},
  \citenamefont {Flouquet}, \citenamefont {Tokunaga},\ and\ \citenamefont
  {Aoki}}]{Miyake2019}%
  \BibitemOpen
  \bibfield  {author} {\bibinfo {author} {\bibfnamefont {A.}~\bibnamefont
  {Miyake}}, \bibinfo {author} {\bibfnamefont {Y.}~\bibnamefont {Shimizu}},
  \bibinfo {author} {\bibfnamefont {Y.~J.}\ \bibnamefont {Sato}}, \bibinfo
  {author} {\bibfnamefont {D.}~\bibnamefont {Li}}, \bibinfo {author}
  {\bibfnamefont {A.}~\bibnamefont {Nakamura}}, \bibinfo {author}
  {\bibfnamefont {Y.}~\bibnamefont {Homma}}, \bibinfo {author} {\bibfnamefont
  {F.}~\bibnamefont {Honda}}, \bibinfo {author} {\bibfnamefont
  {J.}~\bibnamefont {Flouquet}}, \bibinfo {author} {\bibfnamefont
  {M.}~\bibnamefont {Tokunaga}},\ and\ \bibinfo {author} {\bibfnamefont
  {D.}~\bibnamefont {Aoki}},\ }\bibfield  {title} {\bibinfo {title}
  {Metamagnetic transition in heavy fermion superconductor {UTe}$_2$},\ }\href
  {https://doi.org/10.7566/"J. Phys. Soc. Jpn.".88.063706} {\bibfield
  {journal} {\bibinfo  {journal} {Journal of the Physical Society of Japan}\
  }\textbf {\bibinfo {volume} {88}},\ \bibinfo {pages} {063706} (\bibinfo
  {year} {2019})}\BibitemShut {NoStop}%
\bibitem [{\citenamefont {Zacharias}\ and\ \citenamefont
  {Garst}(2013)}]{Zacharias13}%
  \BibitemOpen
  \bibfield  {author} {\bibinfo {author} {\bibfnamefont {M.}~\bibnamefont
  {Zacharias}}\ and\ \bibinfo {author} {\bibfnamefont {M.}~\bibnamefont
  {Garst}},\ }\bibfield  {title} {\bibinfo {title} {Quantum criticality in
  itinerant metamagnets},\ }\href {https://doi.org/10.1103/PhysRevB.87.075119}
  {\bibfield  {journal} {\bibinfo  {journal} {Phys. Rev. B}\ }\textbf {\bibinfo
  {volume} {87}},\ \bibinfo {pages} {075119} (\bibinfo {year}
  {2013})}\BibitemShut {NoStop}%
\bibitem [{\citenamefont {Tokiwa}\ \emph {et~al.}(2013)\citenamefont {Tokiwa},
  \citenamefont {Garst}, \citenamefont {Gegenwart}, \citenamefont {Bud'ko},\
  and\ \citenamefont {Canfield}}]{Tokiwa2013}%
  \BibitemOpen
  \bibfield  {author} {\bibinfo {author} {\bibfnamefont {Y.}~\bibnamefont
  {Tokiwa}}, \bibinfo {author} {\bibfnamefont {M.}~\bibnamefont {Garst}},
  \bibinfo {author} {\bibfnamefont {P.}~\bibnamefont {Gegenwart}}, \bibinfo
  {author} {\bibfnamefont {S.~L.}\ \bibnamefont {Bud'ko}},\ and\ \bibinfo
  {author} {\bibfnamefont {P.~C.}\ \bibnamefont {Canfield}},\ }\bibfield
  {title} {\bibinfo {title} {Quantum bicriticality in the heavy-fermion
  metamagnet {Y}b{A}g{G}e},\ }\href
  {https://doi.org/10.1103/PhysRevLett.111.116401} {\bibfield  {journal}
  {\bibinfo  {journal} {Phys. Rev. Lett.}\ }\textbf {\bibinfo {volume} {111}},\
  \bibinfo {pages} {116401} (\bibinfo {year} {2013})}\BibitemShut {NoStop}%
\bibitem [{\citenamefont {Rost}\ \emph {et~al.}(2009)\citenamefont {Rost},
  \citenamefont {Perry}, \citenamefont {Mercure}, \citenamefont {Mackenzie},\
  and\ \citenamefont {Grigera}}]{Rost-Science09}%
  \BibitemOpen
  \bibfield  {author} {\bibinfo {author} {\bibfnamefont {A.~W.}\ \bibnamefont
  {Rost}}, \bibinfo {author} {\bibfnamefont {R.~S.}\ \bibnamefont {Perry}},
  \bibinfo {author} {\bibfnamefont {J.-F.}\ \bibnamefont {Mercure}}, \bibinfo
  {author} {\bibfnamefont {A.~P.}\ \bibnamefont {Mackenzie}},\ and\ \bibinfo
  {author} {\bibfnamefont {S.~A.}\ \bibnamefont {Grigera}},\ }\bibfield
  {title} {\bibinfo {title} {Entropy landscape of phase formation associated
  with quantum criticality in {S}r$_3${R}u$_2${O}$_7$},\ }\href
  {https://doi.org/10.1126/science.1176627} {\bibfield  {journal} {\bibinfo
  {journal} {Science}\ }\textbf {\bibinfo {volume} {325}},\ \bibinfo {pages}
  {1360} (\bibinfo {year} {2009})}\BibitemShut {NoStop}%
\bibitem [{\citenamefont {Yip}\ \emph {et~al.}(1991)\citenamefont {Yip},
  \citenamefont {Li},\ and\ \citenamefont {Kumar}}]{Yip1991}%
  \BibitemOpen
  \bibfield  {author} {\bibinfo {author} {\bibfnamefont {S.~K.}\ \bibnamefont
  {Yip}}, \bibinfo {author} {\bibfnamefont {T.}~\bibnamefont {Li}},\ and\
  \bibinfo {author} {\bibfnamefont {P.}~\bibnamefont {Kumar}},\ }\bibfield
  {title} {\bibinfo {title} {Thermodynamic considerations and the phase diagram
  of superconducting ${\mathrm{upt}}_{3}$},\ }\href
  {https://doi.org/10.1103/PhysRevB.43.2742} {\bibfield  {journal} {\bibinfo
  {journal} {Phys. Rev. B}\ }\textbf {\bibinfo {volume} {43}},\ \bibinfo
  {pages} {2742} (\bibinfo {year} {1991})}\BibitemShut {NoStop}%
\bibitem [{\citenamefont {Aoki}\ \emph
  {et~al.}(2022{\natexlab{c}})\citenamefont {Aoki}, \citenamefont {Brison},
  \citenamefont {Flouquet}, \citenamefont {Ishida}, \citenamefont {Knebel},
  \citenamefont {Tokunaga},\ and\ \citenamefont {Yanase}}]{aoki2022b}%
  \BibitemOpen
  \bibfield  {author} {\bibinfo {author} {\bibfnamefont {D.}~\bibnamefont
  {Aoki}}, \bibinfo {author} {\bibfnamefont {J.~P.}\ \bibnamefont {Brison}},
  \bibinfo {author} {\bibfnamefont {J.}~\bibnamefont {Flouquet}}, \bibinfo
  {author} {\bibfnamefont {K.}~\bibnamefont {Ishida}}, \bibinfo {author}
  {\bibfnamefont {G.}~\bibnamefont {Knebel}}, \bibinfo {author} {\bibfnamefont
  {Y.}~\bibnamefont {Tokunaga}},\ and\ \bibinfo {author} {\bibfnamefont
  {Y.}~\bibnamefont {Yanase}},\ }\bibfield  {title} {\bibinfo {title}
  {Unconventional superconductivity in {UTe}$_2$},\ }\href
  {https://doi.org/10.1088/1361-648X/ac5863} {\bibfield  {journal} {\bibinfo
  {journal} {J. Phys.: Condens. Matter}\ }\textbf {\bibinfo {volume} {34}},\
  \bibinfo {pages} {243002} (\bibinfo {year} {2022}{\natexlab{c}})}\BibitemShut
  {NoStop}%
\bibitem [{\citenamefont {Fujibayashi}\ \emph {et~al.}(2022)\citenamefont
  {Fujibayashi}, \citenamefont {Nakamine}, \citenamefont {Kinjo}, \citenamefont
  {Kitagawa}, \citenamefont {Ishida}, \citenamefont {Tokunaga}, \citenamefont
  {Sakai}, \citenamefont {Kambe}, \citenamefont {Nakamura}, \citenamefont
  {Shimizu}, \citenamefont {Homma}, \citenamefont {Li}, \citenamefont {Honda},\
  and\ \citenamefont {Aoki}}]{Fujibayashi2022}%
  \BibitemOpen
  \bibfield  {author} {\bibinfo {author} {\bibfnamefont {H.}~\bibnamefont
  {Fujibayashi}}, \bibinfo {author} {\bibfnamefont {G.}~\bibnamefont
  {Nakamine}}, \bibinfo {author} {\bibfnamefont {K.}~\bibnamefont {Kinjo}},
  \bibinfo {author} {\bibfnamefont {S.}~\bibnamefont {Kitagawa}}, \bibinfo
  {author} {\bibfnamefont {K.}~\bibnamefont {Ishida}}, \bibinfo {author}
  {\bibfnamefont {Y.}~\bibnamefont {Tokunaga}}, \bibinfo {author}
  {\bibfnamefont {H.}~\bibnamefont {Sakai}}, \bibinfo {author} {\bibfnamefont
  {S.}~\bibnamefont {Kambe}}, \bibinfo {author} {\bibfnamefont
  {A.}~\bibnamefont {Nakamura}}, \bibinfo {author} {\bibfnamefont
  {Y.}~\bibnamefont {Shimizu}}, \bibinfo {author} {\bibfnamefont
  {Y.}~\bibnamefont {Homma}}, \bibinfo {author} {\bibfnamefont
  {D.}~\bibnamefont {Li}}, \bibinfo {author} {\bibfnamefont {F.}~\bibnamefont
  {Honda}},\ and\ \bibinfo {author} {\bibfnamefont {D.}~\bibnamefont {Aoki}},\
  }\bibfield  {title} {\bibinfo {title} {Superconducting order parameter in
  {UTe}$_{2}$ determined by knight shift measurement},\ }\href
  {https://doi.org/10.7566/"J. Phys. Soc. Jpn.".91.043705} {\bibfield
  {journal} {\bibinfo  {journal} {Journal of the Physical Society of Japan}\
  }\textbf {\bibinfo {volume} {91}},\ \bibinfo {pages} {043705} (\bibinfo
  {year} {2022})}\BibitemShut {NoStop}%


\end{thebibliography}
\end{document}